\documentclass[ALICE,manyauthors]{cernphprep}
\usepackage[comma,square,numbers,sort&compress]{natbib}
\usepackage{hyperref}
\usepackage{lineno}
\usepackage{etoolbox}
\usepackage{xspace}
\usepackage{xcolor}
\usepackage{booktabs}
\usepackage{soul}
\usepackage[T1]{fontenc}
\usepackage{orcidlink}

\begin{document}
%%%%%%%%%%%%%%%%%%%%%%%%%%%%%%%%%%%%%%%%%%%%%%%%%%
% These are some new commands that may be useful 
% for paper writing in general. If other newcommands
% are needed for your specific paper, please feel 
% free to add here. 
%
% The currently available commands are organized in: 
% 1) Systems
% 2) Quantities
% 3) Energies and units
% 4) Detectors
% 5) particle species 
%%%%%%%%%%%%%%%%%%%%%%%%%%%%%%%%%%%%%%%%%%%%%%%%%%

% 1) SYSTEMS 
\newcommand{\pp}           {pp\xspace}
\newcommand{\ppbar}        {\mbox{$\mathrm {p\overline{p}}$}\xspace}
\newcommand{\XeXe}         {\mbox{Xe--Xe}\xspace}
\newcommand{\PbPb}         {\mbox{Pb--Pb}\xspace}
\newcommand{\pA}           {\mbox{pA}\xspace}
\newcommand{\pPb}          {\mbox{p--Pb}\xspace}
\newcommand{\AuAu}         {\mbox{Au--Au}\xspace}
\newcommand{\dAu}          {\mbox{d--Au}\xspace}

% 2) QUANTITIES 
\newcommand{\s}            {\ensuremath{\sqrt{s}}\xspace}
\newcommand{\snn}          {\ensuremath{\sqrt{s_{\mathrm{NN}}}}\xspace}
\newcommand{\pt}           {\ensuremath{p_{\rm T}}\xspace}
\newcommand{\meanpt}       {$\langle p_{\mathrm{T}}\rangle$\xspace}
\newcommand{\ycms}         {\ensuremath{y_{\rm CMS}}\xspace}
\newcommand{\ylab}         {\ensuremath{y_{\rm lab}}\xspace}
\newcommand{\etarange}[1]  {\mbox{$\left | \eta \right |~<~#1$}}
\newcommand{\yrange}[1]    {\mbox{$\left | y \right |~<~#1$}}
\newcommand{\dndy}         {\ensuremath{\mathrm{d}N_\mathrm{ch}/\mathrm{d}y}\xspace}
\newcommand{\dndeta}       {\ensuremath{\mathrm{d}N_\mathrm{ch}/\mathrm{d}\eta}\xspace}
\newcommand{\avdndeta}     {\ensuremath{\langle\dndeta\rangle}\xspace}
\newcommand{\dNdy}         {\ensuremath{\mathrm{d}N_\mathrm{ch}/\mathrm{d}y}\xspace}
\newcommand{\Npart}        {\ensuremath{N_\mathrm{part}}\xspace}
\newcommand{\Ncoll}        {\ensuremath{N_\mathrm{coll}}\xspace}
\newcommand{\dEdx}         {\ensuremath{\textrm{d}E/\textrm{d}x}\xspace}
\newcommand{\RpPb}         {\ensuremath{R_{\rm pPb}}\xspace}

% 3) ENERGIES, UNITS
\newcommand{\nineH}        {$\sqrt{s}~=~0.9$~Te\kern-.1emV\xspace}
\newcommand{\seven}        {$\sqrt{s}~=~7$~Te\kern-.1emV\xspace}
\newcommand{\twoH}         {$\sqrt{s}~=~0.2$~Te\kern-.1emV\xspace}
\newcommand{\twosevensix}  {$\sqrt{s}~=~2.76$~Te\kern-.1emV\xspace}
\newcommand{\five}         {$\sqrt{s}~=~5.02$~Te\kern-.1emV\xspace}
\newcommand{\twosevensixnn}{$\sqrt{s_{\mathrm{NN}}}~=~2.76$~Te\kern-.1emV\xspace}
\newcommand{\fivenn}       {$\sqrt{s_{\mathrm{NN}}}~=~5.02$~Te\kern-.1emV\xspace}
\newcommand{\LT}           {L{\'e}vy-Tsallis\xspace}
\newcommand{\GeVc}         {Ge\kern-.1emV/$c$\xspace}
\newcommand{\MeVc}         {Me\kern-.1emV/$c$\xspace}
\newcommand{\TeV}          {Te\kern-.1emV\xspace}
\newcommand{\GeV}          {Ge\kern-.1emV\xspace}
\newcommand{\MeV}          {Me\kern-.1emV\xspace}
\newcommand{\GeVmass}      {Ge\kern-.2emV/$c^2$\xspace}
\newcommand{\MeVmass}      {Me\kern-.2emV/$c^2$\xspace}
\newcommand{\lumi}         {\ensuremath{\mathcal{L}}\xspace}

% 4) DETECTORS 
\newcommand{\ITS}          {\rm{ITS}\xspace}
\newcommand{\TOF}          {\rm{TOF}\xspace}
\newcommand{\ZDC}          {\rm{ZDC}\xspace}
\newcommand{\ZDCs}         {\rm{ZDCs}\xspace}
\newcommand{\ZNA}          {\rm{ZNA}\xspace}
\newcommand{\ZNC}          {\rm{ZNC}\xspace}
\newcommand{\SPD}          {\rm{SPD}\xspace}
\newcommand{\SDD}          {\rm{SDD}\xspace}
\newcommand{\SSD}          {\rm{SSD}\xspace}
\newcommand{\TPC}          {\rm{TPC}\xspace}
\newcommand{\TRD}          {\rm{TRD}\xspace}
\newcommand{\VZERO}        {\rm{V0}\xspace}
\newcommand{\VZEROA}       {\rm{V0A}\xspace}
\newcommand{\VZEROC}       {\rm{V0C}\xspace}
\newcommand{\Vdecay} 	   {\ensuremath{V^{0}}\xspace}

% 4) PARTICLE SPECIES 
\newcommand{\ee}           {\ensuremath{e^{+}e^{-}}} 
\newcommand{\pip}          {\ensuremath{\pi^{+}}\xspace}
\newcommand{\pim}          {\ensuremath{\pi^{-}}\xspace}
\newcommand{\kap}          {\ensuremath{\rm{K}^{+}}\xspace}
\newcommand{\kam}          {\ensuremath{\rm{K}^{-}}\xspace}
\newcommand{\pbar}         {\ensuremath{\rm\overline{p}}\xspace}
\newcommand{\kzero}        {\ensuremath{{\rm K}^{0}_{\rm{S}}}\xspace}
\newcommand{\lmb}          {\ensuremath{\Lambda}\xspace}
\newcommand{\almb}         {\ensuremath{\overline{\Lambda}}\xspace}
\newcommand{\Om}           {\ensuremath{\Omega^-}\xspace}
\newcommand{\Mo}           {\ensuremath{\overline{\Omega}^+}\xspace}
\newcommand{\X}            {\ensuremath{\Xi^-}\xspace}
\newcommand{\Ix}           {\ensuremath{\overline{\Xi}^+}\xspace}
\newcommand{\Xis}          {\ensuremath{\Xi^{\pm}}\xspace}
\newcommand{\Oms}          {\ensuremath{\Omega^{\pm}}\xspace}
\newcommand{\degree}       {\ensuremath{^{\rm o}}\xspace}

% 5) EXTRA
%\newcommand{\pt}{\ensuremath{p_{\mathrm{T}}}\xspace}
%\newcommand{\mt}{\ensuremath{m_{\mathrm{T}}}\xspace}
\newcommand{\ptAnd}[1]{\ensuremath{p_{\mathrm{T},#1}}\xspace}
\newcommand{\Mjj}{\ensuremath{M_{\mathrm{jj}}}\xspace}
\newcommand{\kt}{\ensuremath{k_{\mathrm{T}}}\xspace}
\newcommand{\antikt}{anti-\kt}
\newcommand{\abs}[1]{\left|#1\right|}
\newcommand{\dd}{\mathrm{d}\xspace}
\newcommand{\RpA}{\ensuremath{R_\mathrm{pA}}\xspace}

\newcommand*\linenomathpatch[1]{%
  \cspreto{#1}{\linenomath}%
  \cspreto{#1*}{\linenomath}%
  \csappto{end#1}{\endlinenomath}%
  \csappto{end#1*}{\endlinenomath}%
}
\linenomathpatch{equation}
\linenomathpatch{align}

%%%%%%%%%%%%%%%  Title page %%%%%%%%%%%%%%%%%%%%%%%%
\begin{titlepage}
% the dates below correspond to CERN approval
% please don't touch: EB chairs will take care
\PHyear{2026}       % required, will be obtained from CERN
\PHnumber{089}      % required, will be obtained from CERN
\PHdate{19 March}  % required, will be obtained from CERN
%%%%%%%%%%%%%%%%%%%%%%%%%%%%%%%%%%%%%%%%%%%%%%%%%%%%

%%% Put your own title + short title here:
    \title{Dijet invariant mass of charged-particle jets in pp and p--Pb collisions at $\mathbf{\sqrt{s_\mathrm{NN}}=5.02}$ TeV}
\ShortTitle{Dijet invariant mass of charged-particle jets in pp and p--Pb}   % appears on left page headers

%%% Do not change the next lines
\Collaboration{ALICE Collaboration\thanks{See Appendix~\ref{app:collab} for the list of collaboration members}}
\ShortAuthor{ALICE Collaboration} % appears on right page headers, do not change

\begin{abstract}
    The ALICE collaboration presents the first measurement of the dijet invariant mass spectra of charged-particle jets in pp and p--Pb collisions at $\sqrt{s_\mathrm{NN}}=5.02$ TeV. Charged particles in the mid-pseudorapidity region, $\abs{\eta}<0.9$, are clustered into jets using the anti-$k_\mathrm{T}$ algorithm with a resolution parameter $R=0.4$. The leading and subleading jets are required to have a transverse momentum $p_\mathrm{T} > 20$~GeV$/c$ and to be contained within $\left|\eta_\mathrm{jet}\right|<0.5$. 
    The dijet invariant mass spectrum and the nuclear modification factor $R_\mathrm{pA}$ are presented in the low-mass region of 75 to 150~GeV/$c^2$. The nuclear modification factor for charged-particle dijet invariant mass is consistent with unity. This is in line with previous small-system jet studies. Comparisons with Monte Carlo simulations suggest that the low-mass region is sensitive to anti-shadowing effects on parton densities in the nucleus, however, the expected signal is subtle and below the present experimental sensitivity.
\end{abstract}
\end{titlepage}

\setcounter{page}{2} %please do not remove this line

%%%%%%%%%%%%%%%%%%%%%%%%%%%%%%%%
% begin main text
%%%%%%%%%%%%%%%%%%%%%%%%%%%%%%%%

{
\section{Introduction} 

Jets are formed from the fragmentation of hard-scattered partons (quarks and gluons), characterized by a large momentum transfer. 
As these processes are well described by perturbative calculations, jets provide valuable tools for testing predictions of quantum chromodynamics (QCD).

In heavy-ion collisions, hard partons lose their energy while they traverse the hot and dense partonic matter, known as the quark--gluon plasma (QGP), created in the collision. This energy loss is typically studied via a suppression of the yield of high-$p_{\rm T}$ particles and jets in heavy-ion collisions~\cite{Bjorken:1982tu,Lebedev:1990un,Thoma:1990fm,Burgess:1991wc,dEnterria:2009xfs}  compared to the number of binary nucleon--nucleon collision scaled yield in pp collisions~\cite{ATLAS:2012tjt,ALICE:2013dpt,ALICE:2018vuu,CMS:2015ved,CMS:2016xef}. Unlike inclusive jet measurements, which are largely dominated by jets emitted near the surface of the medium and thus biased toward survivors of minimal energy loss, dijet systems offer a qualitatively different probe of the QGP, as their correlated topology increases sensitivity to jets that traverse denser regions of the fireball~\cite{Renk:2006pk}. The ATLAS~\cite{Aad:2010bu,ATLAS:2022zbu} and CMS~\cite{CMS:2014qvs} collaborations measured a clear increase of momentum imbalance in dijet systems in heavy-ion collisions compared to pp collisions. 

Energy loss in the QGP is not the only source for modifications of jet yields in heavy-ion collisions. Cold nuclear matter (CNM) effects have been studied extensively, both experimentally~\cite{ALICE:2018vuu,CMS:2015ved,CMS:2016xef,ATLAS:2022kqu} and by identifying various theoretical expectations ranging from modification of the parton densities in heavy nuclei, described with nuclear parton distribution functions (nPDFs)~\cite{Eskola:2021nhw,Kovarik:2015cma,AbdulKhalek:2020yuc}, to gluon saturation effects~\cite{Mueller:1985wy}, or energy loss in CNM~\cite{Arleo:2014oha}. Final states of very high multiplicity pp or p--Pb events may be dense enough such that a small droplet of QGP is formed~\cite{ALICE:2012eyl,CMS:2010ifv,ATLAS:2015hzw}, although recent measurements already give stringent limits for energy losses in small systems~\cite{ALICE:2017svf,CMS:2014qvs}. We investigate CNM where QGP formation is not expected. CNM effects can be quantified using p--Pb collisions with minimum bias (MB), which provide a baseline for Pb--Pb collision measurements. 

In the leading order of perturbative QCD theory, dijets originate from $2\rightarrow2$ processes. Due to momentum conservation, the two outgoing partons of the hard interaction have transverse momenta ($p_\mathrm{T}$) that are opposite and equal in magnitude, $\vec{p}_{{\rm T},1}=-\vec{p}_{{\rm T},2}$. This momentum balance is broken by higher-order processes, like hard $2\rightarrow3$ scatterings~\cite{Frixione:2002ik}, QCD evolution of the incoming partons~\cite{ALICE:2015ppz}, or multiple interactions~\cite{Sjostrand:1985vv}. Hence, it is required that the pair of jets forming the dijet system is measured in opposite hemispheres in the transverse plane.

The invariant mass of the two jets forming the dijet system is called the ``dijet mass" for the rest of this paper. The dijet mass provides a Lorentz-invariant measure for the hardness, or scale $Q$, of the hard interaction. The dijet mass has been studied at very high masses by the ATLAS~\cite{Aad:2011aj} and CMS~\cite{Chatrchyan:2011ns} collaborations, since an unexpected resonance peak in the distributions would be a signature of physics beyond the Standard Model~\cite{CMS:2010qze,CDF:2008ieg,ATLAS:2020zzb}. ALICE is well-positioned to study the dijet mass at a lower mass range, corresponding to lower transverse momentum dijet pairs. In this low-momentum region in heavy-ion collisions, the medium-induced modifications are expected to be more pronounced.

Dijet production measurements in p--Pb collisions have been used to constrain nPDFs~\cite{CMS:2024idr}. In particular, it has been shown that these measurements can provide further evidence of gluon nuclear shadowing at a small momentum fraction $x$, and anti-shadowing at values around $x\approx0.1$~\cite{Eskola:2019dui,Eskola:2021nhw}.
The dijet mass measurements presented in this paper may serve as a reference for future Pb--Pb measurements. 

This paper presents the first measurement of the mass spectrum of charged-particle dijets reconstructed from tracks measured by ALICE in pp and p--Pb collisions at center-of-mass energy $\snn=5.02$ TeV. The nuclear modification factor ($R_{\rm pPb}$) is determined for dijet masses of 75--150 GeV/$c^2$ and compared to Monte Carlo (MC) model calculations. Based on the MC models studied, the low-mass region might be sensitive to anti-shadowing in nPDFs.

\section{Experimental setup and data samples} 
\label{sec:exp}

The ALICE detector is optimised for measuring ultrarelativistic heavy-ion collisions, but it also has a rich pp and p--Pb physics program~\cite{Aamodt:2008zz,ALICE:2022wpn}. This analysis uses data collected during the LHC Run 2 at $\snn~=~5.02$~TeV. The analyzed p--Pb data sample corresponds to an integrated luminosity of $\mathcal{L}_{\text{p-Pb}}=(349\pm 13)~\mu\mathrm{b}^{-1}$ ($730\times 10^6$ MB events)~\cite{Abelev:2014epa}, and the pp data sample to $\mathcal{L}_{\text{pp}}=(20.3\pm 0.5)~\mathrm{nb}^{-1}$ ($1042\times 10^6$ MB events)~\cite{ALICE-PUBLIC-2018-014}. In p--Pb collisions, the center-of-mass is shifted in rapidity by $\Delta y_{\rm cm}=-0.465$, which is towards the direction of the proton beam, because, as the LHC beams are operated at equal beam rigidity, the energy per nucleon of the lead beam is smaller in comparison to the proton beam by a factor Z/A. This analysis utilizes tracking detectors in the ALICE central barrel for vertex determination and reconstruction of tracks, which are used in jet reconstruction, and forward detectors for triggering. 

The data was collected with the ALICE MB trigger based on signals measured with two forward detectors V0A and V0C, located on both sides of the interaction point~\cite{Cortese:2004aa}. The V0A covers forward pseudorapidity $2.8<\eta<5.1$ and V0C backward pseudorapidity $-3.7<\eta<-1.7$. The MB trigger fires when a simultaneous signal in V0A and V0C is seen in coincidence with a bunch crossing~\cite{Aamodt:2008zz}. Further event selection is performed offline to reject events where multiple collisions take place at the same bunch crossing (pileup), by requiring a single primary vertex to be reconstructed, and background events from beam-gas interactions~\cite{ALICE:2014sbx}. These MB events define the visible cross section $\sigma_{\rm V0AND}$ that is measured in van der Meer scans~\cite{vanderMeer:1968zz}.

The ALICE Inner Tracking System (ITS)~\cite{ALICE:1999cls} in Run 2 consists of six layers of silicon detectors. The ITS is used for vertex reconstruction and particle tracking, especially at low $p_\mathrm{T}$. Its two innermost layers form the silicon pixel detector (SPD) that provides accurate vertexing.
The ITS is surrounded by the large Time Projection Chamber (TPC)~\cite{Alme_2010} in a uniform 0.5~T magnetic field along the beam axis. In the ALICE global track reconstruction, TPC standalone tracks are prolonged into the ITS. After ITS points are added to the track, the track parameters are refitted. 

The best momentum resolution is reached with tracks that contain at least one point in the SPD. However, while the TPC-only tracks have an excellent uniformity in azimuth, the $\varphi$--acceptance of tracks containing a hit in the SPD is non-uniform due to detector inefficiencies~\cite{ALICE:2014sbx}. If a track candidate does not contain a hit in the SPD, a primary vertex is included in the track fit. Accepting both kinds of tracks in this analysis allows for a uniform acceptance, without detrimental loss in momentum resolution or a large fake track rate. The overall tracking efficiency of the ALICE experiment is high, with the TPC tracking efficiency ranging from 70\% to 85\% for tracks over 0.5~GeV$/c$ with very little dependency on detector occupancy~\cite{ALICE:2014sbx}. 

The primary vertex is required to be located within $|z|<10$~cm from the nominal interaction point in the beam direction to ensure a symmetric tracking acceptance. 
Events that are triggered but for which no primary vertex is reconstructed are rejected from the analysis. In order to determine the total number of events with or without a vertex inside the fiducial cut $|z|<10$~cm, a vertex reconstruction efficiency $\epsilon_{\rm vtx}$ is calculated. The $\epsilon_{\rm vtx}$  is defined as the number of single collision events with a vertex divided by the number of single collision events with or without a vertex.

Finite tracking efficiencies and other instrumental effects in this analysis are estimated and corrected with detector performance simulations. The particles used to simulate the detector performance during the p--Pb collisions were simulated using the PYTHIA6 MC event generator~\cite{Sjostrand:2006za} with the Perugia 2011 tune~\cite{Skands:2010ak}, while the particles for the detector performance simulations for the pp collisions were simulated using the PYTHIA8 MC event generator~\cite{Bierlich:2022pfr} with the Monash 2013 tune~\cite{Skands:2014pea}. The simulated particles were propagated through the detector material using GEANT3~\cite{Brun:1994aa}, giving access to simulated detector signals which undergo the same event reconstruction and tracking steps as the real signal. The dijet mass analysis is performed both on charged hadrons directly from the MC (MC truth) and charged tracks obtained from the full detector simulation chain (reconstructed at the detector level). A comparison of these results enables the building of response matrices, which are used to unfold the measured spectra from real data. 

To interpret the data and present model comparisons, two event generators are used: PYTHIA8 using the Monash 2013 tune, and a combination of POWHEG and PYTHIA8, where the POWHEG BOX generator~\cite{Nason:2004rx,Frixione:2007vw,Alioli:2010xd} creates a partonic event based on next-to-leading order (NLO) pQCD processes, which is then evolved and hadronized using PYTHIA8. In this simulation, the PYTHIA shower model is retuned to be compatible with NLO cross sections~\cite{Lonnblad:2012ix}. PYTHIA was run in several bins of the transverse momentum of the hard parton scattering, as was done for the purpose of detector performance simulation. On the other hand, POWHEG generates hard events at an enhanced rate, which is then accounted for by scaling with the inverse of the enhancement factor event-by-event.

As a pp baseline, PYTHIA and POWHEG+PYTHIA were run using CT14NLO~\cite{Dulat:2015mca} and CT18ANLO~\cite{Hou:2019efy} proton PDFs, respectively. To study the cold nuclear matter effects from nPDFs, the simulations were repeated such that the positive rapidity going proton PDF was modified with nuclear modifications in lead. In the case of PYTHIA,  this was done using EPPS16~\cite{Eskola:2016oht}, and in the case of POWHEG+PYTHIA using EPPS21~\cite{Eskola:2021nhw}. The PDF sets and nPDF modifications need to be selected consistently because the analysis of nuclear modifications is built on some chosen free proton parton distributions. These simulations take into account the rapidity shift of the center-of-mass system $\Delta y=-0.465$ in p--Pb collisions. For PYTHIA, this was achieved by defining the energies of the beams asymmetrically, and for POWHEG+PYTHIA, the whole event was boosted by the $\Delta y$.

\section{Data analysis} 
\subsection{Jet reconstruction}
Jets are reconstructed from charged-particle tracks using the FastJet package~\cite{Cacciari:2011ma} with the \antikt algorithm~\cite{Cacciari:2008gp} with a resolution parameter of $R=0.4$, for tracks with $p_\mathrm{T} > 150~\mathrm{MeV}/c$. The clustering uses the $p_\mathrm{T}$ recombination scheme~\cite{Cacciari:2011ma} in which reconstructed tracks are treated as massless pseudo-jets and combined such that the resulting jet four-momentum is massless by construction. The $p_\mathrm{T}$ scheme is optimised to reproduce the transverse momentum of the jet when there is no particle identification (PID) information, or it is limited. As the mass of a single jet is not crucial in this analysis, the method is adopted to gain more straightforward comparability between data and theory models, as the uncertainties related to the limited PID do not play a role in the measurement. 

Jets are required to be contained within the fiducial acceptance of the TPC, $\left|\eta_\mathrm{jet}\right|<0.5$. The underlying event consisting of soft interactions is subtracted from the reconstructed jet clusters using the four-momentum background subtraction~\cite{Cacciari:2007fd,Soyez:2012hv}
\begin{equation}
        p_{\mathrm{corr}}^\mu=p^\mu - \left[\left(\rho +\rho_m \right)A_\mathrm{jet}^E,\ \rho A_\mathrm{jet}^x,\ \rho A_\mathrm{jet}^y,\ \left(\rho +\rho_m \right)A_\mathrm{jet}^z\right],
\end{equation}
where $A_\mathrm{jet}$ is the jet area four-vector~\cite{Cacciari:2011ma} in the $\varphi$--$\eta$ plane calculated using the explicit ghost method~\cite{Cacciari:2011ma}, $\rho$ is the \pt density of the event, and $\rho_m$ is the mass density of the jets. To estimate the background density, jet clusters reconstructed using the \kt~algorithm~\cite{Ellis:1993tq} are adopted, and for each event $\rho$ is calculated using the sparse event method for both pp and p--Pb,
\begin{equation} \label{eq:rho}
    \rho=\underset{\mathrm{real}\ \kt\ \mathrm{jets}}{\mathrm{median}}\left\lbrace\frac{\ptAnd{\mathrm{jet}}}{A_\mathrm{jet}} \right\rbrace \times C,
\end{equation}
where the ``real \kt jets'' refer to \kt jets with at least one track, i.e., pure ghost jets are not included in calculations of the median. Also, the leading and subleading jets are dropped from the determination of $\rho$. In this analysis, $\rho_m$ is always zero, as the \pt scheme clusters jets to be massless. The empty area of the event is taken into account with the coefficient 
\begin{equation}
    C=\frac{\sum_{\mathrm{real}\ \kt\ \mathrm{jets}}A_j}{\sum_{\mathrm{all}\ \kt\ \mathrm{jets}}A_j},
\end{equation}
which estimates the empty $\varphi$--$\eta$ area of the event using jets clustered with the \kt algorithm. This technique is described in detail in reference~\cite{Adam:2015hoa}.

Dijets are considered in events where the transverse momenta of both the leading ($p_{\rm{T},1}$) and subleading ($p_{\rm{T},2}$) jets are greater than $20$~GeV$/c$ and the jets are in opposite hemispheres in the transverse plane, i.e., have an azimuthal difference $|\Delta\varphi|>\pi/2$. This requirement is applied in order to suppress dijets in which the leading and subleading jets originate from different hard scatterings. The dijet invariant mass is then calculated as
\begin{align}
     \Mjj^2&= m_1^2 + m_2^2 + 2 \left( m_\mathrm{T,1} m_\mathrm{T,2} \cosh( \Delta y ) - \ptAnd{1}\ptAnd{2} \cos( \Delta\varphi ) \right)\\
            &\approx 2\ptAnd{1} \ptAnd{2} \left( \cosh(\Delta\eta) - \cos(\Delta\varphi) \right),
\end{align}
where $m_i$ and $m_\mathrm{T,i}$ are the mass and transverse mass of the leading and subleading jets. The last approximation is exact for jets with a massless four-momentum. In this analysis, we consider only the two hardest (largest $p_\mathrm{T}$) jets in the event.

\subsection{Background fluctuations}
The dijet mass is calculated from jets after background subtraction. However, the background at the location of the jet might be different from the overall event density determined by Eq.~(\ref{eq:rho}), and hence the dijet mass distributions will be affected by background fluctuations. Similarly to earlier ALICE analysis in \PbPb~\cite{ALICE:2013dpt}, these fluctuations are estimated by measuring a local background density using rotated cones with $R=0.4$, and calculating the subtracted dijet mass with the local background. This new result, which includes the fluctuations, is compared to the default method determined by Eq.~(\ref{eq:rho}),
\begin{equation}
    \delta \Mjj \equiv \Mjj - \Mjj',
\end{equation}
where \Mjj is the invariant mass calculated using the default description of the background density $\rho$, and $\Mjj'$ is calculated using a background density $\rho'$ estimated with rotated cones. The two rotated cones are created such that the first cone is perpendicular in the azimuthal plane to the leading jet, and the second cone is perpendicular to the subleading jet. Both cones are rotated in the same direction so that the rotated cones have the same $\Delta\varphi$ as the original dijet. The background density is calculated separately for both leading and subleading jets from transverse momentum density $\rho' \equiv \Sigma_\mathrm{cone} \ptAnd{i} / \left( \pi R^2\right)$ inside the rotated cones. 

In the case of an overlap between a cone and a leading or subleading jet, the event is not considered in the background determination. 
On rare occasions, the rotated cone overlaps with another hard jet in the event, which results in a significant and long tail at $\delta\Mjj>0$, as $\Mjj'$ becomes small due to overcorrecting the background. In such conditions, $\delta\Mjj$ does not measure true background fluctuations, and all events which result in a negative $\Mjj'$ in the analysis are rejected from the $\delta\Mjj$ distribution. After the rejection, the tail in the $\delta\Mjj$ distribution is largely suppressed.
 
The probability distributions for $\delta\Mjj$ fluctuations in pp and p--Pb are shown in Fig.~\ref{fig:deltaM}. It is seen that the $\delta\Mjj$ distribution is peaked close to zero but asymmetric around the mean $\mu\approx0$. Hence, the distribution is fitted with an asymmetric generalized Gaussian distribution allowing different width parameters ($\sigma_\pm$) and powers ($q^\pm\ne2$) for positive ($x-\mu\ge0$) and negative sides of the mean:
\begin{equation}
    f(\delta M_\mathrm{jj})=C\times\left\{ 
    \begin{array}{ll}
        &\exp{\left[-\left(\frac{\abs{x-\mu}}{\sigma^+}\right)^{q^+}\right]}, \hspace{1em} x-\mu \ge 0 \\
        \vspace{-0.8em} \\
        &\exp{\left[-\left(\frac{\abs{x-\mu}}{\sigma^-}\right)^{q^-}\right]}, \hspace{1em} x-\mu < 0
    \end{array}
    \right. .
\end{equation}
It is important to note that the right (left) width, i.e., $|x-\mu|\ge0$ ($<0$), of the distribution depends both on $q^\pm$ and $\sigma^\pm$. As the tails of the distribution do not describe fluctuations, the fit range is restricted to $-10<\delta\Mjj<35$~GeV$/c^2$. We expect the p--Pb distribution to exhibit larger fluctuations, therefore the power  $q^+$ has been restricted so that the p--Pb peak ends up wider than the pp peak.

In what follows, only a sufficient description of the shape of the $\delta\Mjj$ distribution is needed. The chosen shape of the fit function does not have any particular physical interpretation. In practice, the background fluctuations are treated together with unfolding of the detector effects. Later, the fit of the $\delta\Mjj$ distribution is used in the unfolding procedure instead of the measured distribution itself to avoid effects from statistical fluctuations in unfolding.

\begin{figure}%
    \begin{center}
    \includegraphics[width = 0.70\textwidth]{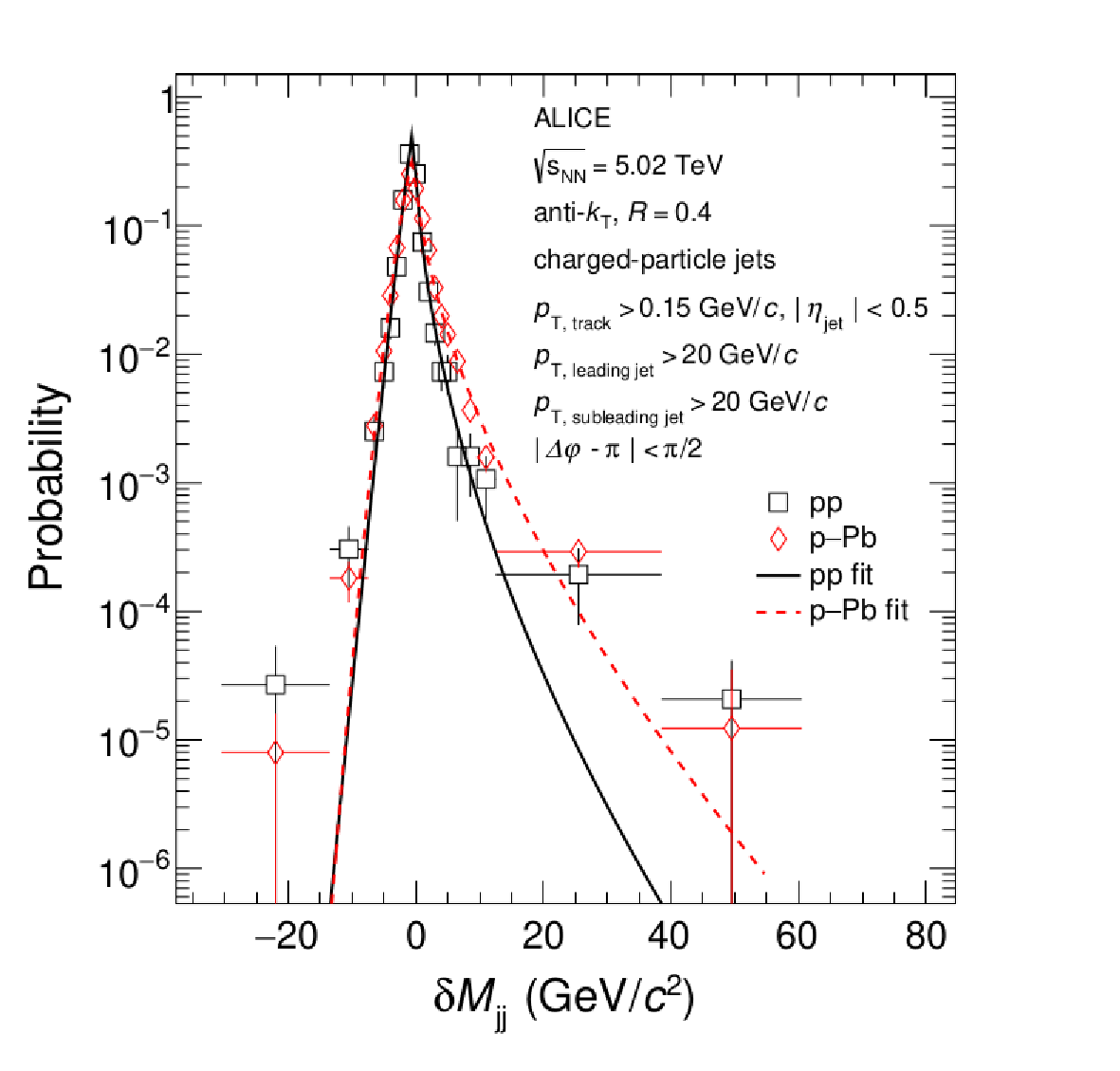}
    \end{center}
    \caption{The probability distribution of the background fluctuation in pp and p--Pb collisions fitted with an asymmetric generalized Gaussian.}
    \label{fig:deltaM}
\end{figure}

\subsection{Unfolding}
In order to account for background fluctuations and instrumental effects, an unfolding procedure is used to translate the measured dijet mass distribution into the physical cross section. A response matrix is constructed using a MC simulation as described in Sec.~\ref{sec:exp}. The same analysis was performed for detector level MC data as was done for the real data. Simulated jets in the true and detector level are matched geometrically by requiring that the distance in the $(\varphi,\eta)$ plane $\Delta R<0.3$. The truth level jets consist of final state charged particles with a lifetime $\tau > 1 ~\mathrm{cm/}c$~\cite{ALICE:2017hcy}. When a match for both leading and subleading jets is found, an entry is filled to the unfolding matrix $\hat{M}_{\rm det}$. A dijet is considered to be matched even if the detector level leading jet is matched to the truth level subleading jet and vice versa. In the case of a true dijet not being matched with the reconstructed jets, this event is counted as a miss. Conversely, if an event contains a reconstructed fake dijet, this is counted as a fake. It may happen that a single event contains a missed true dijet and a reconstructed fake. The fake rate is below~5\% within the range of the measurement. While there are~$\sim30$\% misses at the lowest mass bin, the fraction reduces to~$\sim15$\% at high masses.

The unfolding matrix $\hat{M}_{\rm det}$ includes only instrumental detector effects. Methods of a previous analysis~\cite{Adam:2016jfp,Adam:2015hoa,ALICE:2019qyj} are followed in order to include background fluctuations into the unfolding procedure. A matrix $\hat{M}_{\rm fluct}$ is constructed to describe fluctuations measured at the detector level, such that every column vector of the matrix follows the fit of the $\delta\Mjj$-probability distribution presented in Fig.~\ref{fig:deltaM}. This construction preserves the number of dijets in the smearing with the fluctuations based on standard matrix multiplication rules. It is assumed that the background fluctuations do not depend strongly on the hardness of the event. This was checked by studying the $\delta\Mjj$ distributions in $\Mjj$ bins, and it is found that the assumption holds well within statistical uncertainties in the region of small $|\delta\Mjj|<20$~GeV$/c^2$ associated with fluctuations. The total unfolding matrix can be obtained with matrix multiplication:
\begin{equation}
\hat{M}_{\rm tot}=\hat{M}_{\rm fluct}\hat{M}_{\rm det}.
\label{eq:unf}
\end{equation}

The unfolding is performed using a Bayesian iterative method with three iterations for the p--Pb spectrum and five iterations for the pp spectrum using the RooUnfold package~\cite{Adye:2011gm}. After the unfolding, each bin of the dijet spectrum is corrected with the fraction of the missed true and reconstructed fake dijets that were found during the matching process in that particular bin. 

\subsection{Observables}
The differential cross section for the dijet mass is defined as
\begin{equation} \label{eq:xsec}
    \frac{\dd\sigma}{\dd M_\mathrm{jj}}=\frac{\sigma_\mathrm{MB}\epsilon_\mathrm{vtx}}{N_\mathrm{events}}\frac{\Delta N}{\Delta M_\mathrm{jj}},
\end{equation}
where $\sigma_\mathrm{MB}$ is the visible cross section corresponding to the ALICE minimum bias trigger, $\epsilon_\mathrm{vtx}$ is the vertex reconstruction efficiency, $\Delta N$ is the fully corrected number of dijets in a given dijet mass bin, and $\Delta \Mjj$ is the bin width. The dijets are reconstructed from jets within the fiducial acceptance, $\left|\eta_\mathrm{jet}\right|<0.5$. The visible cross sections are measured in van der Meer scans: in p--Pb collisions $\sigma^\mathrm{pPb}_\mathrm{MB}=(2090 \pm 70\ ({\rm syst.}))$~mb~\cite{Abelev:2014epa} and in pp collisions $\sigma^\mathrm{pp}_\mathrm{MB}=(50.87 \pm 0.04 \ ({\rm stat.}) \pm 0.92 \ ({\rm syst.}) )$~mb~\cite{ALICE-PUBLIC-2018-014}. The vertex reconstruction efficiencies are $\epsilon_\mathrm{vtx}^\mathrm{pPb}=0.985$, and $\epsilon_\mathrm{vtx}^\mathrm{pp}=0.949$. Finally, $N_{\rm events}$ is the total number of events selected by the MB trigger in the data sample with vertex. For the minimum bias collisions, the nuclear modification factor is defined as
\begin{equation} \label{eq:nuclear_mod}
    R_\mathrm{pPb}=\left.\frac{\dd \sigma^\mathrm{pPb}}{\dd\Mjj} \right/ \left(\mathrm{A}\frac{\dd \sigma^\mathrm{pp}}{\dd\Mjj}\hspace{0.4em}\right),
\end{equation}
where the mass number of the lead nucleus $A=208$.

\section{Systematic uncertainties}

The systematic uncertainties of the measurement are dominated by the tracking efficiency uncertainty that enters through the unfolding process, where the masses of the true and detector level dijets are matched. Due to finite tracking efficiency, a fraction of particles are not correctly reconstructed and hence the values of reconstructed $p_{\rm T,jet}$ are smaller than the true transverse momenta. As the tracking efficiency is known with a finite precision, this uncertainty is inherited by the unfolding systematic uncertainty.

The uncertainty of the tracking efficiency is estimated to be a constant 3\% for pp collisions, while for p--Pb the uncertainty varies as a function of track \pt from 1\% at under 2 GeV/$c$ to 2.5\% at 8 GeV/$c$ and above. To estimate the effects of this uncertainty, reconstructed tracks are randomly removed at a rate according to the corresponding uncertainty. After removing the tracks, the jets are re-reconstructed and a modified unfolding matrix $\hat{M}_{\rm det}$ is constructed and used in the unfolding process instead of the default matrix. Observed differences in the final dijet mass distributions are assigned as the tracking efficiency uncertainty. This procedure leads to a larger unfolding correction because the number of particles available for jet reconstruction is reduced in the detector level MC. This corresponds to a situation where the true tracking efficiency would be smaller than what the current detector simulations give. It is expected that the effect is symmetric.

The performance of the unfolding is validated in several ways. The overall performance of the unfolding is validated with a closure test, where a MC sample that is statistically independent of that used to determine the response matrix is unfolded, and the result is compared to the MC truth in that sample. The effects of using a different truncation of the histograms and a different choice of binning during the unfolding are studied. The sensitivity of the results to the change of the unfolding prior is also studied. A prior is the initial guess of the shape of the unfolded product, so the prior is changed to the shape of the unfolded data, while the default prior is the MC source used in the detector response calculation. Also, a different number of iterations of the Bayesian unfolding is considered, and the dependence on the choice of the unfolding method is studied by repeating the analysis with an unfolding technique leveraging singular value decomposition (SVD)~\cite{Hocker:1995kb}. The regularization value of the SVD unfolding process is, by default, selected to be the square of the $k^\text{th}$ singular value. For this analysis, the number $k$ has been selected from the $d$-vector to be 9 for the pp spectrum and 5 for the p--Pb spectrum.

Following reference~\cite{ALICE:2022vsz}, the observed differences are combined as the total systematic uncertainty of unfolding as
\begin{equation}\delta_\mathrm{unfolding}=\sqrt{\frac{\sum^N_{i}\delta_i^2}{N}},
\end{equation}
where the sum is over all the unfolding systematic uncertainties. This is done since each of the unfolding systematics comprises independent measurements of the same underlying systematic uncertainty. The final uncertainties are calculated bin-by-bin, and the results are smoothed separately for the tracking and the combined unfolding uncertainties before combining them. The tracking and unfolding uncertainties are added in quadrature. 

The tracking efficiency uncertainty contributes 7--8\% across the spectrum for p--Pb and 10--13\% for pp. The unfolding systematic uncertainty adds 0--0.5\% for p--Pb and 0--2\% for pp. As precise correlations in the tracking systematic uncertainty between different collision systems are not known, we follow a conservative estimate by taking the maximum deviation between mixed default/modified ratios as the point‑by‑point systematic uncertainty of~\RpPb.

\section{Results and Conclusions}

The first ALICE measurement of the dijet invariant mass cross section (Eq.~(\ref{eq:xsec})) reconstructed from charged tracks at central rapidity is shown in the top panel of Fig.~\ref{fig:final}. The pp dijet cross section, scaled by the mass number of the lead ion ($A=208$), is compared with the p--Pb dijet cross section.  The bottom panel presents the nuclear modification factor \RpPb, defined in Eq.~(\ref{eq:nuclear_mod}). The \RpPb is found to agree with unity within uncertainties. 

\begin{figure}[tb]
    \begin{center}
    \includegraphics[width = 0.70\textwidth]{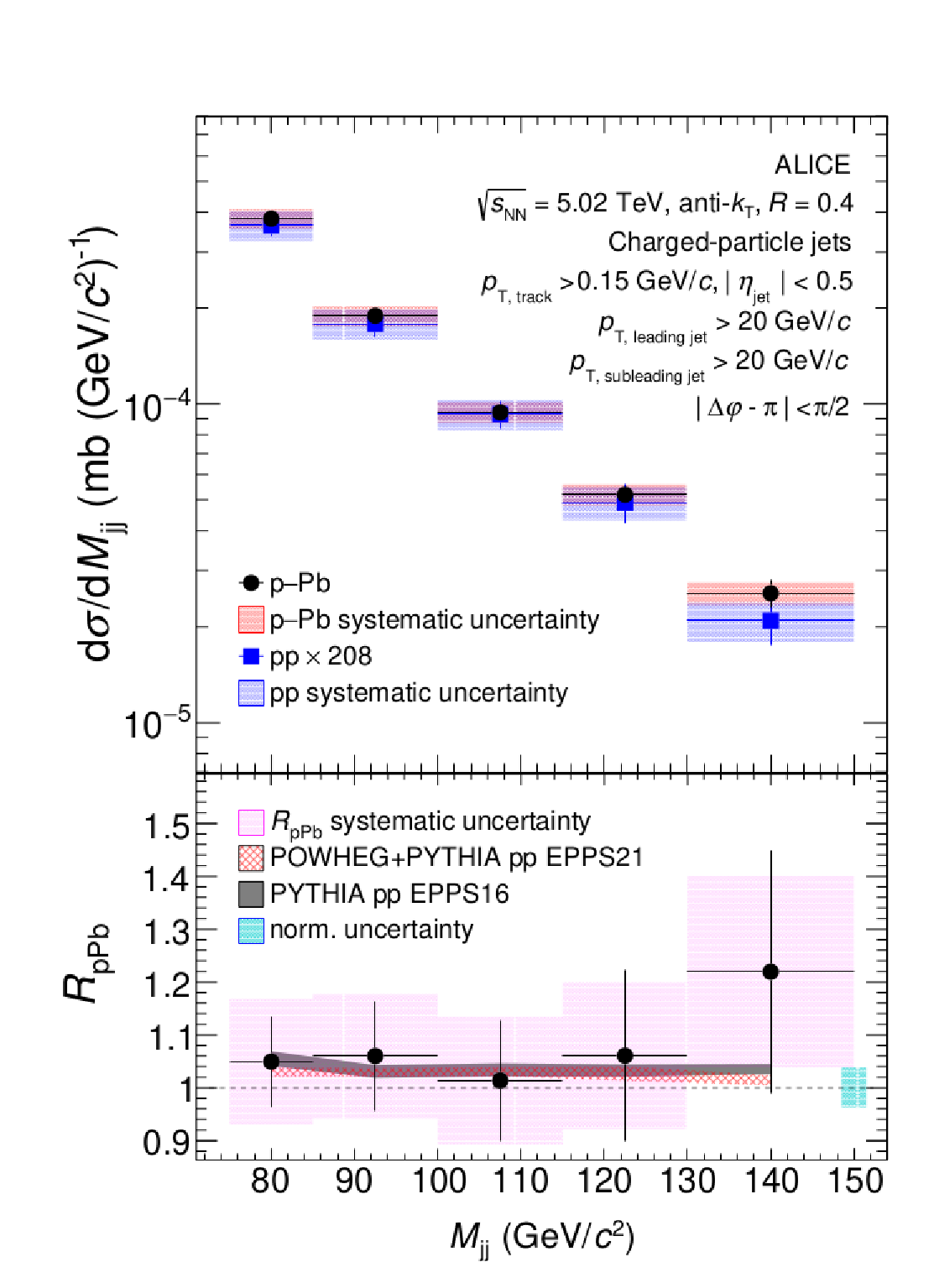}
    \end{center}
    \caption{The dijet invariant mass spectrum in pp and p--Pb at \fivenn is presented in the top panel, and the nuclear modification factor \RpPb in the bottom panel. In the bottom panel, the simulation results from PYTHIA and POWHEG+PYTHIA are also shown.}
    \label{fig:final}
\end{figure}

To understand the observed trends and better interpret the results, the results are compared with MC simulations. PYTHIA and POWHEG+PYTHIA were used to generate collisions that use proton PDFs to simulate pp collisions and a combination of proton PDFs and lead nPDFs with a rapidity boost to generate p--Pb collisions, as explained in section~\ref{sec:exp}.
The \RpPb for the simulations was determined by dividing the simulated dijet mass cross section, including nPDFs, by the baseline pp cross section. Figure~\ref{fig:final} shows the simulation results with grey and hashed red bands, reflecting the statistical uncertainties of the simulations. Both simulations exhibit a slight enhancement of 1--7 percent of the dijet yield in p--Pb and are in agreement with the measured data. During the simulation, it was checked what momentum fractions $(x_1,x_2)$ the partons entering the hardest QCD interaction have in events that contain a dijet with the selections posed in this analysis. Both fractions turn out to be close to the anti-shadowing region in nPDF's around $x\sim0.01$, and, based on this, the slight enhancement is attributed to a hint of anti-shadowing. However, the expected effect is small, and the relatively large uncertainties in the data limit its discriminating power.

The first ALICE measurement for the dijet mass cross sections of jets reconstructed from charged tracks at central rapidity is presented for proton-proton and proton-lead collisions down to small masses. It is found that the corresponding nuclear modification factor $\RpPb$ is consistent with unity, and no significant cold nuclear matter effects are observed within the current experimental uncertainties. 
Simulations, in which nPDFs were used to modify one of the colliding protons, suggest that dijets in this kinematic range could probe the anti-shadowing region in nPDFs, as the momentum fractions of partons in the hardest QCD interaction are close to the anti-shadowing region. The effect, however, is subtle, and the simulations are consistent with the data within uncertainties. Future measurements with Run 4 data can improve the sensitivity of the measurement due to a significant increase in recorded luminosity, with the projected integrated luminosity during a one-month p--Pb run projected to be around 300 $\rm{nb}^{-1}$~\cite{Bruce:2722753}. These results provide a reference for corresponding measurements in Pb--Pb collisions. 
}

%%%%% acknowledgements - handled by EB chairs 
\newenvironment{acknowledgement}{\relax}{\relax}
\begin{acknowledgement}
\section*{Acknowledgements}
% add specific acknowledgements here 
% ...but please don't remove the line below: funding agencies
% will be acknowledged with a custom tex file handled by EB chairs after Collab Round 2
% Version: 2026-03-03

The ALICE Collaboration would like to thank all its engineers and technicians for their invaluable contributions to the construction of the experiment and the CERN accelerator teams for the outstanding performance of the LHC complex.
The ALICE Collaboration gratefully acknowledges the resources and support provided by all Grid centres and the Worldwide LHC Computing Grid (WLCG) collaboration.
The ALICE Collaboration acknowledges the following funding agencies for their support in building and running the ALICE detector:
A. I. Alikhanyan National Science Laboratory (Yerevan Physics Institute) Foundation (ANSL), State Committee of Science and World Federation of Scientists (WFS), Armenia;
Austrian Academy of Sciences, Austrian Science Fund (FWF): [M 2467-N36] and Nationalstiftung f\"{u}r Forschung, Technologie und Entwicklung, Austria;
Ministry of Communications and High Technologies, National Nuclear Research Center, Azerbaijan;
Rede Nacional de Física de Altas Energias (Renafae), Financiadora de Estudos e Projetos (Finep), Funda\c{c}\~{a}o de Amparo \`{a} Pesquisa do Estado de S\~{a}o Paulo (FAPESP) and The Sao Paulo Research Foundation  (FAPESP), Brazil;
Bulgarian Ministry of Education and Science, within the National Roadmap for Research Infrastructures 2020-2027 (object CERN), Bulgaria;
Ministry of Education of China (MOEC) , Ministry of Science \& Technology of China (MSTC) and National Natural Science Foundation of China (NSFC), China;
Ministry of Science and Education and Croatian Science Foundation, Croatia;
Centro de Aplicaciones Tecnol\'{o}gicas y Desarrollo Nuclear (CEADEN), Cubaenerg\'{\i}a, Cuba;
Ministry of Education, Youth and Sports of the Czech Republic, Czech Republic;
The Danish Council for Independent Research | Natural Sciences, the VILLUM FONDEN and Danish National Research Foundation (DNRF), Denmark;
Helsinki Institute of Physics (HIP), Finland;
Commissariat \`{a} l'Energie Atomique (CEA) and Institut National de Physique Nucl\'{e}aire et de Physique des Particules (IN2P3) and Centre National de la Recherche Scientifique (CNRS), France;
Bundesministerium f\"{u}r Forschung, Technologie und Raumfahrt (BMFTR) and GSI Helmholtzzentrum f\"{u}r Schwerionenforschung GmbH, Germany;
National Research, Development and Innovation Office, Hungary;
Department of Atomic Energy Government of India (DAE), Department of Science and Technology, Government of India (DST), University Grants Commission, Government of India (UGC) and Council of Scientific and Industrial Research (CSIR), India;
National Research and Innovation Agency - BRIN, Indonesia;
Istituto Nazionale di Fisica Nucleare (INFN), Italy;
Japanese Ministry of Education, Culture, Sports, Science and Technology (MEXT) and Japan Society for the Promotion of Science (JSPS) KAKENHI, Japan;
Consejo Nacional de Ciencia (CONACYT) y Tecnolog\'{i}a, through Fondo de Cooperaci\'{o}n Internacional en Ciencia y Tecnolog\'{i}a (FONCICYT) and Direcci\'{o}n General de Asuntos del Personal Academico (DGAPA), Mexico;
Nederlandse Organisatie voor Wetenschappelijk Onderzoek (NWO), Netherlands;
The Research Council of Norway, Norway;
Pontificia Universidad Cat\'{o}lica del Per\'{u}, Peru;
Ministry of Science and Higher Education, National Science Centre and WUT ID-UB, Poland;
Korea Institute of Science and Technology Information and National Research Foundation of Korea (NRF), Republic of Korea;
Ministry of Education and Scientific Research, Institute of Atomic Physics, Ministry of Research and Innovation and Institute of Atomic Physics and Universitatea Nationala de Stiinta si Tehnologie Politehnica Bucuresti, Romania;
Ministerstvo skolstva, vyskumu, vyvoja a mladeze SR, Slovakia;
National Research Foundation of South Africa, South Africa;
Swedish Research Council (VR) and Knut \& Alice Wallenberg Foundation (KAW), Sweden;
European Organization for Nuclear Research, Switzerland;
Suranaree University of Technology (SUT), National Science and Technology Development Agency (NSTDA) and National Science, Research and Innovation Fund (NSRF via PMU-B B05F650021), Thailand;
Turkish Energy, Nuclear and Mineral Research Agency (TENMAK), Turkey;
National Academy of  Sciences of Ukraine, Ukraine;
Science and Technology Facilities Council (STFC), United Kingdom;
National Science Foundation of the United States of America (NSF) and United States Department of Energy, Office of Nuclear Physics (DOE NP), United States of America.
In addition, individual groups or members have received support from:
FORTE project, reg.\ no.\ CZ.02.01.01/00/22\_008/0004632, Czech Republic, co-funded by the European Union, Czech Republic;
European Research Council (grant no. 950692), European Union;
Deutsche Forschungs Gemeinschaft (DFG, German Research Foundation) ``Neutrinos and Dark Matter in Astro- and Particle Physics'' (grant no. SFB 1258), Germany.

\end{acknowledgement}

%%%%%%%% Bibliography 
\bibliographystyle{utphys}   % Remember we use title in the biblio
\bibliography{references.bib}

\providecommand{\href}[2]{#2}\begingroup\raggedright\begin{thebibliography}{10}

\bibitem{Bjorken:1982tu}
J.~D. Bjorken, ``{Energy Loss of Energetic Partons in Quark - Gluon Plasma:
  Possible Extinction of High p(t) Jets in Hadron - Hadron Collisions}'',
  \href{https://inspirehep.net/literature/181746}{FERMILAB-PUB-82-059-THY,
  FERMILAB-PUB-82-059-T}.

\bibitem{Lebedev:1990un}
V.~V. Lebedev and A.~V. Smilga, ``{On anomalous damping in quark - gluon
  plasma}'', \href{https://doi.org/10.1016/0370-2693(91)91389-D}{{\em Phys.
  Lett.} {\bfseries B253} (1991) 231--236}.

\bibitem{Thoma:1990fm}
M.~H. Thoma and M.~Gyulassy, ``{Quark Damping and Energy Loss in the High
  Temperature {QCD}}'',
  \href{https://doi.org/10.1016/S0550-3213(05)80031-8}{{\em Nucl. Phys.}
  {\bfseries B351} (1991) 491--506}.

\bibitem{Burgess:1991wc}
C.~P. Burgess and A.~L. Marini, ``{The Damping of energetic gluons and quarks
  in high temperature QCD}'',
  \href{https://doi.org/10.1103/PhysRevD.45.R17}{{\em Phys. Rev.} {\bfseries
  D45} (1992) 17--20}, \href{https://arxiv.org/abs/hep-th/9109051}{{\ttfamily
  arXiv:hep-th/9109051 [hep-th]}}.

\bibitem{dEnterria:2009xfs}
D.~d'Enterria, ``{Jet quenching}'',
  \href{https://doi.org/10.1007/978-3-642-01539-7_16}{{\em Landolt-Bornstein}
  {\bfseries 23} (2010) 471},
\href{https://arxiv.org/abs/0902.2011}{{\ttfamily arXiv:0902.2011 [nucl-ex]}}.
%%CITATION = ARXIV:0902.2011;%%.

\bibitem{ATLAS:2012tjt}
{\bfseries ATLAS} Collaboration, G.~Aad {\em et~al.}, ``{Measurement of the jet
  radius and transverse momentum dependence of inclusive jet suppression in
  lead-lead collisions at $\sqrt{s_{NN}}$= 2.76 TeV with the ATLAS detector}'',
  \href{https://doi.org/10.1016/j.physletb.2013.01.024}{{\em Phys. Lett.}
  {\bfseries B719} (2013) 220--241},
  \href{https://arxiv.org/abs/1208.1967}{{\ttfamily arXiv:1208.1967 [hep-ex]}}.

\bibitem{ALICE:2013dpt}
{\bfseries ALICE} Collaboration, B.~Abelev {\em et~al.}, ``{Measurement of
  charged jet suppression in Pb-Pb collisions at $\sqrt{s_{NN}}$ = 2.76 TeV}'',
  \href{https://doi.org/10.1007/JHEP03(2014)013}{{\em J. High Energy Phys.}
  {\bfseries 03} (2014) 013}, \href{https://arxiv.org/abs/1311.0633}{{\ttfamily
  arXiv:1311.0633 [nucl-ex]}}.

\bibitem{ALICE:2018vuu}
{\bfseries ALICE} Collaboration, S.~Acharya {\em et~al.}, ``{Transverse
  momentum spectra and nuclear modification factors of charged particles in pp,
  p-Pb and Pb-Pb collisions at the LHC}'',
  \href{https://doi.org/10.1007/JHEP11(2018)013}{{\em J. High Energy Phys.}
  {\bfseries 11} (2018) 013},
  \href{https://arxiv.org/abs/1802.09145}{{\ttfamily arXiv:1802.09145
  [nucl-ex]}}.

\bibitem{CMS:2015ved}
{\bfseries CMS} Collaboration, V.~Khachatryan {\em et~al.}, ``{Nuclear Effects
  on the Transverse Momentum Spectra of Charged Particles in pPb Collisions at
  $\sqrt{s_{_\mathrm {NN}}} =5.02$ TeV}'',
  \href{https://doi.org/10.1140/epjc/s10052-015-3435-4}{{\em Eur. Phys. J.}
  {\bfseries C75} (2015) 237},
  \href{https://arxiv.org/abs/1502.05387}{{\ttfamily arXiv:1502.05387
  [nucl-ex]}}.

\bibitem{CMS:2016xef}
{\bfseries CMS} Collaboration, V.~Khachatryan {\em et~al.}, ``{Charged-particle
  nuclear modification factors in PbPb and pPb collisions at $
  \sqrt{s_{\mathrm{N}\;\mathrm{N}}}=5.02 $ TeV}'',
  \href{https://doi.org/10.1007/JHEP04(2017)039}{{\em J. High Energy Phys.}
  {\bfseries 04} (2017) 039},
  \href{https://arxiv.org/abs/1611.01664}{{\ttfamily arXiv:1611.01664
  [nucl-ex]}}.

\bibitem{Renk:2006pk}
T.~Renk and K.~Eskola, ``{Prospects of medium tomography using back-to-back
  hadron correlations}'',
  \href{https://doi.org/10.1103/PhysRevC.75.054910}{{\em Phys. Rev.} {\bfseries
  C75} (2007) 054910},
\href{https://arxiv.org/abs/hep-ph/0610059}{{\ttfamily arXiv:hep-ph/0610059
  [hep-ph]}}.
%%CITATION = HEP-PH/0610059;%%.

\bibitem{Aad:2010bu}
{\bfseries ATLAS} Collaboration, G.~Aad {\em et~al.}, ``{Observation of a
  Centrality-Dependent Dijet Asymmetry in Lead-Lead Collisions at
  $\sqrt{s_\mathrm{NN}}=2.77$ TeV with the ATLAS Detector at the LHC}'',
  \href{https://doi.org/10.1103/PhysRevLett.105.252303}{{\em Phys. Rev. Lett.}
  {\bfseries 105} (2010) 252303},
\href{https://arxiv.org/abs/1011.6182}{{\ttfamily arXiv:1011.6182 [hep-ex]}}.
%%CITATION = ARXIV:1011.6182;%%.

\bibitem{ATLAS:2022zbu}
{\bfseries ATLAS} Collaboration, G.~Aad {\em et~al.}, ``{Measurements of the
  suppression and correlations of dijets in Pb+Pb collisions at
  $\sqrt{s_{_\text{NN}}}$=5.02 TeV}'',
  \href{https://doi.org/10.1103/PhysRevC.107.054908}{{\em Phys. Rev.}
  {\bfseries C107} (2023) 054908},
  \href{https://arxiv.org/abs/2205.00682}{{\ttfamily arXiv:2205.00682
  [nucl-ex]}}.

\bibitem{CMS:2014qvs}
{\bfseries CMS} Collaboration, S.~Chatrchyan {\em et~al.}, ``{Studies of dijet
  transverse momentum balance and pseudorapidity distributions in pPb
  collisions at $\sqrt{s_{\mathrm{NN}}} = 5.02$ $\,\text {TeV}$}'',
  \href{https://doi.org/10.1140/epjc/s10052-014-2951-y}{{\em Eur. Phys. J.}
  {\bfseries C74} (2014) 2951},
  \href{https://arxiv.org/abs/1401.4433}{{\ttfamily arXiv:1401.4433
  [nucl-ex]}}.

\bibitem{ATLAS:2022kqu}
{\bfseries ATLAS} Collaboration, G.~Aad {\em et~al.}, ``{Charged-hadron
  production in $pp$, $p$+Pb, Pb+Pb, and Xe+Xe collisions at
  $\sqrt{s_{_\text{NN}}}=5$ TeV with the ATLAS detector at the LHC}'',
  \href{https://doi.org/10.1007/JHEP07(2023)074}{{\em J. High Energy Phys.}
  {\bfseries 07} (2023) 074},
  \href{https://arxiv.org/abs/2211.15257}{{\ttfamily arXiv:2211.15257
  [hep-ex]}}.

\bibitem{Eskola:2021nhw}
K.~J. Eskola, P.~Paakkinen, H.~Paukkunen, and C.~A. Salgado, ``{EPPS21: a
  global QCD analysis of nuclear PDFs}'',
  \href{https://doi.org/10.1140/epjc/s10052-022-10359-0}{{\em Eur. Phys. J.}
  {\bfseries C82} (2022) 413},
  \href{https://arxiv.org/abs/2112.12462}{{\ttfamily arXiv:2112.12462
  [hep-ph]}}.

\bibitem{Kovarik:2015cma}
K.~Kovarik {\em et~al.}, ``{nCTEQ15 - Global analysis of nuclear parton
  distributions with uncertainties in the CTEQ framework}'',
  \href{https://doi.org/10.1103/PhysRevD.93.085037}{{\em Phys. Rev.} {\bfseries
  D93} (2016) 085037}, \href{https://arxiv.org/abs/1509.00792}{{\ttfamily
  arXiv:1509.00792 [hep-ph]}}.

\bibitem{AbdulKhalek:2020yuc}
R.~Abdul~Khalek, J.~J. Ethier, J.~Rojo, and G.~van Weelden, ``{nNNPDF2.0: quark
  flavor separation in nuclei from LHC data}'',
  \href{https://doi.org/10.1007/JHEP09(2020)183}{{\em J. High Energy Phys.}
  {\bfseries 09} (2020) 183},
  \href{https://arxiv.org/abs/2006.14629}{{\ttfamily arXiv:2006.14629
  [hep-ph]}}.

\bibitem{Mueller:1985wy}
A.~H. Mueller and J.-w. Qiu, ``{Gluon Recombination and Shadowing at Small
  Values of x}'', \href{https://doi.org/10.1016/0550-3213(86)90164-1}{{\em
  Nucl. Phys.} {\bfseries B268} (1986) 427--452}.

\bibitem{Arleo:2014oha}
F.~Arleo and S.~Peign\'e, ``{Quarkonium suppression in heavy-ion collisions
  from coherent energy loss in cold nuclear matter}'',
  \href{https://doi.org/10.1007/JHEP10(2014)073}{{\em J. High Energy Phys.}
  {\bfseries 10} (2014) 073}, \href{https://arxiv.org/abs/1407.5054}{{\ttfamily
  arXiv:1407.5054 [hep-ph]}}.

\bibitem{ALICE:2012eyl}
{\bfseries ALICE} Collaboration, B.~Abelev {\em et~al.}, ``{Long-range angular
  correlations on the near and away side in $p$-Pb collisions at
  $\sqrt{s_{NN}}=5.02$ TeV}'',
  \href{https://doi.org/10.1016/j.physletb.2013.01.012}{{\em Phys. Lett.}
  {\bfseries B719} (2013) 29--41},
  \href{https://arxiv.org/abs/1212.2001}{{\ttfamily arXiv:1212.2001
  [nucl-ex]}}.

\bibitem{CMS:2010ifv}
{\bfseries CMS} Collaboration, V.~Khachatryan {\em et~al.}, ``{Observation of
  Long-Range Near-Side Angular Correlations in Proton-Proton Collisions at the
  LHC}'', \href{https://doi.org/10.1007/JHEP09(2010)091}{{\em J. High Energy
  Phys.} {\bfseries 09} (2010) 091},
  \href{https://arxiv.org/abs/1009.4122}{{\ttfamily arXiv:1009.4122 [hep-ex]}}.

\bibitem{ATLAS:2015hzw}
{\bfseries ATLAS} Collaboration, G.~Aad {\em et~al.}, ``{Observation of
  Long-Range Elliptic Azimuthal Anisotropies in $\sqrt{s}=$13 and 2.76 TeV $pp$
  Collisions with the ATLAS Detector}'',
  \href{https://doi.org/10.1103/PhysRevLett.116.172301}{{\em Phys. Rev. Lett.}
  {\bfseries 116} (2016) 172301},
  \href{https://arxiv.org/abs/1509.04776}{{\ttfamily arXiv:1509.04776
  [hep-ex]}}.

\bibitem{ALICE:2017svf}
{\bfseries ALICE} Collaboration, S.~Acharya {\em et~al.}, ``{Constraints on jet
  quenching in p-Pb collisions at $\mathbf{\sqrt{s_{NN}}}$ = 5.02 TeV measured
  by the event-activity dependence of semi-inclusive hadron-jet
  distributions}'', \href{https://doi.org/10.1016/j.physletb.2018.05.059}{{\em
  Phys. Lett.} {\bfseries B783} (2018) 95--113},
  \href{https://arxiv.org/abs/1712.05603}{{\ttfamily arXiv:1712.05603
  [nucl-ex]}}.

\bibitem{Frixione:2002ik}
S.~Frixione and B.~R. Webber, ``{Matching NLO QCD computations and parton
  shower simulations}'',
  \href{https://doi.org/10.1088/1126-6708/2002/06/029}{{\em J. High Energy
  Phys.} {\bfseries 06} (2002) 029},
  \href{https://arxiv.org/abs/hep-ph/0204244}{{\ttfamily arXiv:hep-ph/0204244
  [hep-ph]}}.

\bibitem{ALICE:2015ppz}
{\bfseries ALICE} Collaboration, J.~Adam {\em et~al.}, ``{Measurement of dijet
  $k_T$ in p{\textendash}Pb collisions at $\sqrt{s}_{NN}$=5.02 TeV}'',
  \href{https://doi.org/10.1016/j.physletb.2015.05.033}{{\em Phys. Lett. B}
  {\bfseries 746} (2015) 385--395},
  \href{https://arxiv.org/abs/1503.03050}{{\ttfamily arXiv:1503.03050
  [nucl-ex]}}.

\bibitem{Sjostrand:1985vv}
T.~Sjostrand, ``{Multiple Parton-Parton Interactions in Hadronic Events}'', in
  {\em {23rd International Conference on High-Energy Physics}}.
\newblock 8, 1985.
\newblock \href{https://cds.cern.ch/record/162427}{FERMILAB-PUB-85-119-T}.

\bibitem{Aad:2011aj}
{\bfseries ATLAS} Collaboration, G.~Aad {\em et~al.}, ``{Search for New Physics
  in Dijet Mass and Angular Distributions in pp Collisions at $\sqrt{s} = 7$
  TeV Measured with the ATLAS Detector}'',
  \href{https://doi.org/10.1088/1367-2630/13/5/053044}{{\em New J. Phys.}
  {\bfseries 13} (2011) 053044},
\href{https://arxiv.org/abs/1103.3864}{{\ttfamily arXiv:1103.3864 [hep-ex]}}.
%%CITATION = ARXIV:1103.3864;%%.

\bibitem{Chatrchyan:2011ns}
{\bfseries CMS} Collaboration, S.~Chatrchyan {\em et~al.}, ``{Search for
  Resonances in the Dijet Mass Spectrum from 7 TeV pp Collisions at CMS}'',
  \href{https://doi.org/10.1016/j.physletb.2011.09.015}{{\em Phys. Lett.}
  {\bfseries B704} (2011) 123--142},
\href{https://arxiv.org/abs/1107.4771}{{\ttfamily arXiv:1107.4771 [hep-ex]}}.
%%CITATION = ARXIV:1107.4771;%%.

\bibitem{CMS:2010qze}
{\bfseries CMS} Collaboration, V.~Khachatryan {\em et~al.}, ``{Search for Dijet
  Resonances in 7 TeV pp Collisions at CMS}'',
  \href{https://doi.org/10.1103/PhysRevLett.105.211801}{{\em Phys. Rev. Lett.}
  {\bfseries 105} (2010) 211801},
  \href{https://arxiv.org/abs/1010.0203}{{\ttfamily arXiv:1010.0203 [hep-ex]}}.

\bibitem{CDF:2008ieg}
{\bfseries CDF} Collaboration, T.~Aaltonen {\em et~al.}, ``{Search for new
  particles decaying into dijets in proton-antiproton collisions at $\sqrt{s} =
  1.96$ TeV}'', \href{https://doi.org/10.1103/PhysRevD.79.112002}{{\em Phys.
  Rev.} {\bfseries D79} (2009) 112002},
  \href{https://arxiv.org/abs/0812.4036}{{\ttfamily arXiv:0812.4036 [hep-ex]}}.

\bibitem{ATLAS:2020zzb}
{\bfseries ATLAS} Collaboration, G.~Aad {\em et~al.}, ``{Search for dijet
  resonances in events with an isolated charged lepton using $\sqrt{s} = 13$
  TeV proton-proton collision data collected by the ATLAS detector}'',
  \href{https://doi.org/10.1007/JHEP06(2020)151}{{\em J. High Energy Phys.}
  {\bfseries 06} (2020) 151},
  \href{https://arxiv.org/abs/2002.11325}{{\ttfamily arXiv:2002.11325
  [hep-ex]}}.

\bibitem{CMS:2024idr}
{\bfseries CMS} Collaboration, ``{Constraining nPDFs using dijet production in
  pPb collisions at 8.16 TeV with the CMS experiment}'',
  \href{https://cds.cern.ch/record/2917837}{CMS-PAS-HIN-24-014}.

\bibitem{Eskola:2019dui}
K.~J. Eskola, P.~Paakkinen, and H.~Paukkunen, ``{Non-quadratic improved Hessian
  PDF reweighting and application to CMS dijet measurements at 5.02 TeV}'',
  \href{https://doi.org/10.1140/epjc/s10052-019-6982-2}{{\em Eur. Phys. J.}
  {\bfseries C79} (2019) 511},
  \href{https://arxiv.org/abs/1903.09832}{{\ttfamily arXiv:1903.09832
  [hep-ph]}}.

\bibitem{Aamodt:2008zz}
{\bfseries ALICE} Collaboration, K.~Aamodt {\em et~al.}, ``{The ALICE
  experiment at the CERN LHC}'',
\href{https://doi.org/10.1088/1748-0221/3/08/S08002}{{\em J. Instrum.}
  {\bfseries 3} (2008) S08002}.
%%CITATION = JINST,3,S08002;%%.

\bibitem{ALICE:2022wpn}
{\bfseries ALICE} Collaboration, S.~Acharya {\em et~al.}, ``{The ALICE
  experiment: a journey through QCD}'',
  \href{https://doi.org/10.1140/epjc/s10052-024-12935-y}{{\em Eur. Phys. J.}
  {\bfseries C84} (2024) 813},
  \href{https://arxiv.org/abs/2211.04384}{{\ttfamily arXiv:2211.04384
  [nucl-ex]}}.

\bibitem{Abelev:2014epa}
{\bfseries ALICE} Collaboration, B.~B. Abelev {\em et~al.}, ``{Measurement of
  visible cross sections in proton-lead collisions at $\sqrt{s_\mathrm{NN}}$ =
  5.02 TeV in van der Meer scans with the ALICE detector}'',
  \href{https://doi.org/10.1088/1748-0221/9/11/P11003}{{\em J. Instrum.}
  {\bfseries 9} (2014) P11003},
\href{https://arxiv.org/abs/1405.1849}{{\ttfamily arXiv:1405.1849 [nucl-ex]}}.
%%CITATION = ARXIV:1405.1849;%%.

\bibitem{ALICE-PUBLIC-2018-014}
{\bfseries ALICE} Collaboration, ``{ALICE 2017 luminosity determination for pp
  collisions at $\sqrt{s}$ = 5 TeV}'',
  \href{https://cds.cern.ch/record/2648933}{ALICE-PUBLIC-2018-014}.

\bibitem{Cortese:2004aa}
{\bfseries ALICE} Collaboration, P.~Cortese {\em et~al.}, ``{ALICE technical
  design report on forward detectors: FMD, T0 and V0}'',
\href{https://cds.cern.ch/record/781854}{CERN-LHCC-2004-025}.
%%CITATION = CERN-LHCC-2004-025;%%.

\bibitem{ALICE:2014sbx}
{\bfseries ALICE} Collaboration, B.~B. Abelev {\em et~al.}, ``{Performance of
  the ALICE Experiment at the CERN LHC}'',
  \href{https://doi.org/10.1142/S0217751X14300440}{{\em International J. Mod.
  Phys.} {\bfseries A29} (2014) 1430044},
  \href{https://arxiv.org/abs/1402.4476}{{\ttfamily arXiv:1402.4476
  [nucl-ex]}}.

\bibitem{vanderMeer:1968zz}
S.~van~der Meer, ``{Calibration of the Effective Beam Height in the ISR}'',.
\href{https://cds.cern.ch/record/296752}{CERN-ISR-PO-68-31, ISR-PO-68-31}.
%%CITATION = CERN-ISR-PO-68-31;%%.

\bibitem{ALICE:1999cls}
{\bfseries ALICE} Collaboration, G.~Dellacasa {\em et~al.}, ``{ALICE Inner
  Tracking System (ITS) : Technical Design Report}'',.
  \href{https://cds.cern.ch/record/391175}{CERN-LHCC-99-12}.

\bibitem{Alme_2010}
J.~Alme, Y.~Andres, {\em et~al.}, ``The {ALICE} {TPC}, a large 3-dimensional
  tracking device with fast readout for ultra-high multiplicity events'',
  \href{https://doi.org/10.1016/j.nima.2010.04.042}{{\em Nucl. Instrum. Methods
  Phys. Res.} {\bfseries A622} (Oct, 2010) 316--367}.

\bibitem{Sjostrand:2006za}
T.~Sjostrand, S.~Mrenna, and P.~Z. Skands, ``{PYTHIA 6.4 Physics and Manual}'',
  \href{https://doi.org/10.1088/1126-6708/2006/05/026}{{\em J. High Energy
  Phys.} {\bfseries 05} (2006) 026},
  \href{https://arxiv.org/abs/hep-ph/0603175}{{\ttfamily arXiv:hep-ph/0603175
  [hep-ph]}}.

\bibitem{Skands:2010ak}
P.~Z. Skands, ``{Tuning Monte Carlo Generators: The Perugia Tunes}'',
  \href{https://doi.org/10.1103/PhysRevD.82.074018}{{\em Phys. Rev.} {\bfseries
  D82} (2010) 074018}, \href{https://arxiv.org/abs/1005.3457}{{\ttfamily
  arXiv:1005.3457 [hep-ph]}}.

\bibitem{Bierlich:2022pfr}
C.~Bierlich, S.~Chakraborty, {\em et~al.}, ``{A comprehensive guide to the
  physics and usage of PYTHIA 8.3}'',
  \href{https://doi.org/10.21468/SciPostPhysCodeb.8}{{\em SciPost Phys.
  Codebases} (2022) 8}, \href{https://arxiv.org/abs/2203.11601}{{\ttfamily
  arXiv:2203.11601 [hep-ph]}}.

\bibitem{Skands:2014pea}
P.~Skands, S.~Carrazza, and J.~Rojo, ``{Tuning PYTHIA 8.1: the Monash 2013
  Tune}'', \href{https://doi.org/10.1140/epjc/s10052-014-3024-y}{{\em Eur.
  Phys. J.} {\bfseries C74} (2014) 3024},
\href{https://arxiv.org/abs/1404.5630}{{\ttfamily arXiv:1404.5630 [hep-ph]}}.
%%CITATION = ARXIV:1404.5630;%%.

\bibitem{Brun:1994aa}
R.~Brun, F.~Bruyant, F.~Carminati, S.~Giani, M.~Maire, A.~McPherson,
  G.~Patrick, and L.~Urban, ``{GEANT Detector Description and Simulation
  Tool}'',
\href{https://cds.cern.ch/record/1082634}{W5013, W-5013, CERN-W5013,
  CERN-W-5013}.
%%CITATION = CERN-W5013;%%.

\bibitem{Nason:2004rx}
P.~Nason, ``{A New method for combining NLO QCD with shower Monte Carlo
  algorithms}'', \href{https://doi.org/10.1088/1126-6708/2004/11/040}{{\em J.
  High Energy Phys.} {\bfseries 11} (2004) 040},
\href{https://arxiv.org/abs/hep-ph/0409146}{{\ttfamily arXiv:hep-ph/0409146
  [hep-ph]}}.
%%CITATION = HEP-PH/0409146;%%.

\bibitem{Frixione:2007vw}
S.~Frixione, P.~Nason, and C.~Oleari, ``{Matching NLO QCD computations with
  Parton Shower simulations: the POWHEG method}'',
  \href{https://doi.org/10.1088/1126-6708/2007/11/070}{{\em J. High Energy
  Phys.} {\bfseries 11} (2007) 070},
\href{https://arxiv.org/abs/0709.2092}{{\ttfamily arXiv:0709.2092 [hep-ph]}}.
%%CITATION = ARXIV:0709.2092;%%.

\bibitem{Alioli:2010xd}
S.~Alioli, P.~Nason, C.~Oleari, and E.~Re, ``{A general framework for
  implementing NLO calculations in shower Monte Carlo programs: the POWHEG
  BOX}'', \href{https://doi.org/10.1007/JHEP06(2010)043}{{\em J. High Energy
  Phys.} {\bfseries 06} (2010) 043},
\href{https://arxiv.org/abs/1002.2581}{{\ttfamily arXiv:1002.2581 [hep-ph]}}.
%%CITATION = ARXIV:1002.2581;%%.

\bibitem{Lonnblad:2012ix}
L.~L{\"o}nnblad and S.~Prestel, ``{Merging Multi-leg NLO Matrix Elements with
  Parton Showers}'', \href{https://doi.org/10.1007/JHEP03(2013)166}{{\em J.
  High Energy Phys.} {\bfseries 03} (2013) 166},
  \href{https://arxiv.org/abs/1211.7278}{{\ttfamily arXiv:1211.7278 [hep-ph]}}.

\bibitem{Dulat:2015mca}
S.~Dulat, T.-J. Hou, J.~Gao, M.~Guzzi, J.~Huston, P.~Nadolsky, J.~Pumplin,
  C.~Schmidt, D.~Stump, and C.~P. Yuan, ``{New parton distribution functions
  from a global analysis of quantum chromodynamics}'',
  \href{https://doi.org/10.1103/PhysRevD.93.033006}{{\em Phys. Rev.} {\bfseries
  D93} (2016) 033006}, \href{https://arxiv.org/abs/1506.07443}{{\ttfamily
  arXiv:1506.07443 [hep-ph]}}.

\bibitem{Hou:2019efy}
T.-J. Hou {\em et~al.}, ``{New CTEQ global analysis of quantum chromodynamics
  with high-precision data from the LHC}'',
  \href{https://doi.org/10.1103/PhysRevD.103.014013}{{\em Phys. Rev.}
  {\bfseries D103} (2021) 014013},
  \href{https://arxiv.org/abs/1912.10053}{{\ttfamily arXiv:1912.10053
  [hep-ph]}}.

\bibitem{Eskola:2016oht}
K.~J. Eskola, P.~Paakkinen, H.~Paukkunen, and C.~A. Salgado, ``{EPPS16: Nuclear
  parton distributions with LHC data}'',
  \href{https://doi.org/10.1140/epjc/s10052-017-4725-9}{{\em Eur. Phys. J.}
  {\bfseries C77} (2017) 163},
  \href{https://arxiv.org/abs/1612.05741}{{\ttfamily arXiv:1612.05741
  [hep-ph]}}.

\bibitem{Cacciari:2011ma}
M.~Cacciari, G.~P. Salam, and G.~Soyez, ``{FastJet User Manual}'',
  \href{https://doi.org/10.1140/epjc/s10052-012-1896-2}{{\em Eur. Phys. J.}
  {\bfseries C72} (2012) 1896},
\href{https://arxiv.org/abs/1111.6097}{{\ttfamily arXiv:1111.6097 [hep-ph]}}.
%%CITATION = ARXIV:1111.6097;%%.

\bibitem{Cacciari:2008gp}
M.~Cacciari, G.~P. Salam, and G.~Soyez, ``{The Anti-k(t) jet clustering
  algorithm}'', \href{https://doi.org/10.1088/1126-6708/2008/04/063}{{\em J.
  High Energy Phys.} {\bfseries 04} (2008) 063},
\href{https://arxiv.org/abs/0802.1189}{{\ttfamily arXiv:0802.1189 [hep-ph]}}.
%%CITATION = ARXIV:0802.1189;%%.

\bibitem{Cacciari:2007fd}
M.~Cacciari and G.~P. Salam, ``{Pileup subtraction using jet areas}'',
  \href{https://doi.org/10.1016/j.physletb.2007.09.077}{{\em Phys. Lett.}
  {\bfseries B659} (2008) 119--126},
\href{https://arxiv.org/abs/0707.1378}{{\ttfamily arXiv:0707.1378 [hep-ph]}}.
%%CITATION = ARXIV:0707.1378;%%.

\bibitem{Soyez:2012hv}
G.~Soyez, G.~P. Salam, J.~Kim, S.~Dutta, and M.~Cacciari, ``{Pileup subtraction
  for jet shapes}'', \href{https://doi.org/10.1103/PhysRevLett.110.162001}{{\em
  Phys. Rev. Lett.} {\bfseries 110} (2013) 162001},
\href{https://arxiv.org/abs/1211.2811}{{\ttfamily arXiv:1211.2811 [hep-ph]}}.
%%CITATION = ARXIV:1211.2811;%%.

\bibitem{Ellis:1993tq}
S.~D. Ellis and D.~E. Soper, ``{Successive combination jet algorithm for hadron
  collisions}'', \href{https://doi.org/10.1103/PhysRevD.48.3160}{{\em Phys.
  Rev.} {\bfseries D48} (1993) 3160--3166},
  \href{https://arxiv.org/abs/hep-ph/9305266}{{\ttfamily arXiv:hep-ph/9305266
  [hep-ph]}}.

\bibitem{Adam:2015hoa}
{\bfseries ALICE} Collaboration, J.~Adam {\em et~al.}, ``{Measurement of
  charged jet production cross sections and nuclear modification in p-Pb
  collisions at $\sqrt{s_\mathrm{NN}} = 5.02$ TeV}'',
  \href{https://doi.org/10.1016/j.physletb.2015.07.054}{{\em Phys. Lett.}
  {\bfseries B749} (2015) 68--81},
\href{https://arxiv.org/abs/1503.00681}{{\ttfamily arXiv:1503.00681
  [nucl-ex]}}.
%%CITATION = ARXIV:1503.00681;%%.

\bibitem{ALICE:2017hcy}
{\bfseries ALICE} Collaboration, ``{The ALICE definition of primary
  particles}'',
  \href{https://cds.cern.ch/record/2270008}{ALICE-PUBLIC-2017-005}.

\bibitem{Adam:2016jfp}
{\bfseries ALICE} Collaboration, J.~Adam {\em et~al.}, ``{Centrality dependence
  of charged jet production in p–Pb collisions at $\sqrt{s_\mathrm{NN}}$ =
  5.02 TeV}'', \href{https://doi.org/10.1140/epjc/s10052-016-4107-8}{{\em Eur.
  Phys. J.} {\bfseries C76} (2016) 271},
\href{https://arxiv.org/abs/1603.03402}{{\ttfamily arXiv:1603.03402
  [nucl-ex]}}.
%%CITATION = ARXIV:1603.03402;%%.

\bibitem{ALICE:2019qyj}
{\bfseries ALICE} Collaboration, S.~Acharya {\em et~al.}, ``{Measurements of
  inclusive jet spectra in pp and central Pb-Pb collisions at
  $\sqrt{s_{\rm{NN}}}$ = 5.02 TeV}'',
  \href{https://doi.org/10.1103/PhysRevC.101.034911}{{\em Phys. Rev.}
  {\bfseries C101} (2020) 034911},
  \href{https://arxiv.org/abs/1909.09718}{{\ttfamily arXiv:1909.09718
  [nucl-ex]}}.

\bibitem{Adye:2011gm}
T.~Adye, \href{https://doi.org/10.5170/CERN-2011-006.313}{``{Unfolding
  algorithms and tests using RooUnfold}'',} in {\em {Proceedings, PHYSTAT 2011
  Workshop on Statistical Issues Related to Discovery Claims in Search
  Experiments and Unfolding, CERN,Geneva, Switzerland 17-20 January 2011}},
  pp.~313--318, CERN.
\newblock CERN, Geneva, 2011.
\newblock
\href{https://arxiv.org/abs/1105.1160}{{\ttfamily arXiv:1105.1160
  [physics.data-an]}}.
\newblock
%%CITATION = ARXIV:1105.1160;%%.

\bibitem{Hocker:1995kb}
A.~Hocker and V.~Kartvelishvili, ``{SVD approach to data unfolding}'',
  \href{https://doi.org/10.1016/0168-9002(95)01478-0}{{\em Nucl. Instrum.
  Meth.} {\bfseries A372} (1996) 469--481},
\href{https://arxiv.org/abs/hep-ph/9509307}{{\ttfamily arXiv:hep-ph/9509307
  [hep-ph]}}.
%%CITATION = HEP-PH/9509307;%%.

\bibitem{ALICE:2022vsz}
{\bfseries ALICE} Collaboration, S.~Acharya {\em et~al.}, ``{Measurement of
  inclusive and leading subjet fragmentation in pp and Pb\textendash{}Pb
  collisions at $ \sqrt{s_{\textrm{NN}}} $ = 5.02 TeV}'',
  \href{https://doi.org/10.1007/JHEP05(2023)245}{{\em J. High Energy Phys.}
  {\bfseries 05} (2023) 245},
  \href{https://arxiv.org/abs/2204.10270}{{\ttfamily arXiv:2204.10270
  [nucl-ex]}}.

\bibitem{Bruce:2722753}
R.~Bruce, T.~Argyropoulos, {\em et~al.}, ``{HL-LHC operational scenarios for
  Pb-Pb and p-Pb operation}'', tech. rep., CERN, Geneva, 2020.
\newblock \href{https://cds.cern.ch/record/2722753}{CERN-ACC-2020-0011}.

\end{thebibliography}\endgroup
%\input {bibliography.tex}  

%%%%%%%%%%%%%%%%%%%%%%%%%%%%%%%%
% Appendices: yours (if any) + authorlist
%%%%%%%%%%%%%%%%%%%%%%%%%%%%%%%%
\newpage
\appendix

%
%\input{} % put your appendices here (if any)
%

%%%%% Authorlist - please do not touch: handled by EB chairs 
\section{The ALICE Collaboration}
\label{app:collab}
% ALICE Collaboration author list for 2026-03-03
\begin{flushleft} 
\small

D.A.H.~Abdallah\,\orcidlink{0000-0003-4768-2718}\,$^{\rm 134}$, 
I.J.~Abualrob\,\orcidlink{0009-0005-3519-5631}\,$^{\rm 112}$, 
S.~Acharya\,\orcidlink{0000-0002-9213-5329}\,$^{\rm 49}$, 
K.~Agarwal\,\orcidlink{0000-0001-5781-3393}\,$^{\rm II,}$$^{\rm 23}$, 
G.~Aglieri Rinella\,\orcidlink{0000-0002-9611-3696}\,$^{\rm 32}$, 
L.~Aglietta\,\orcidlink{0009-0003-0763-6802}\,$^{\rm 24}$, 
N.~Agrawal\,\orcidlink{0000-0003-0348-9836}\,$^{\rm 25}$, 
Z.~Ahammed\,\orcidlink{0000-0001-5241-7412}\,$^{\rm 132}$, 
S.~Ahmad\,\orcidlink{0000-0003-0497-5705}\,$^{\rm 15}$, 
I.~Ahuja\,\orcidlink{0000-0002-4417-1392}\,$^{\rm 36}$
% ...

\section*{Affiliation Notes}

$^{\rm I}$ Deceased\\
$^{\rm II}$ Also at: INFN Trieste\\
$^{\rm III}$ Also at: Fondazione Bruno Kessler (FBK), Trento, Italy\\
$^{\rm IV}$ Also at: Czech Technical University in Prague (CZ)\\
$^{\rm V}$ Also at: Instituto de Fisica da Universidade de Sao Paulo\\
$^{\rm VI}$ Also at: Dipartimento DET del Politecnico di Torino, Turin, Italy\\
$^{\rm VII}$ Also at: Department of Applied Physics, Aligarh Muslim University, Aligarh, India\\
$^{\rm VIII}$ Also at: Institute of Theoretical Physics, University of Wroclaw, Poland\\
$^{\rm IX}$ Also at: Facultad de Ciencias, Universidad Nacional Aut\'{o}noma de M\'{e}xico, Mexico City, Mexico\\

\section*{Collaboration Institutes}

$^{1}$ A.I. Alikhanyan National Science Laboratory (Yerevan Physics Institute) Foundation, Yerevan, Armenia\\
$^{2}$ AGH University of Krakow, Cracow, Poland\\
$^{3}$ Bogolyubov Institute for Theoretical Physics, National Academy of Sciences of Ukraine, Kyiv, Ukraine\\
$^{4}$ Bose Institute, Department of Physics  and Centre for Astroparticle Physics and Space Science (CAPSS), Kolkata, India\\
$^{5}$ California Polytechnic State University, San Luis Obispo, California, United States\\
$^{6}$ Central China Normal University, Wuhan, China\\
$^{7}$ Centro de Aplicaciones Tecnol\'{o}gicas y Desarrollo Nuclear (CEADEN), Havana, Cuba\\
$^{8}$ Centro de Investigaci\'{o}n y de Estudios Avanzados (CINVESTAV), Mexico City and M\'{e}rida, Mexico\\
$^{9}$ Chicago State University, Chicago, Illinois, United States\\
$^{10}$ China Nuclear Data Center, China Institute of Atomic Energy, Beijing, China\\
$^{11}$ China University of Geosciences, Wuhan, China\\
$^{12}$ Chungbuk National University, Cheongju, Republic of Korea\\
$^{13}$ Comenius University Bratislava, Faculty of Mathematics, Physics and Informatics, Bratislava, Slovak Republic\\
$^{14}$ Creighton University, Omaha, Nebraska, United States\\
$^{15}$ Department of Physics, Aligarh Muslim University, Aligarh, India\\
$^{16}$ Department of Physics, Pusan National University, Pusan, Republic of Korea\\
$^{17}$ Department of Physics, Sejong University, Seoul, Republic of Korea\\
$^{18}$ Department of Physics, University of California, Berkeley, California, United States\\
$^{19}$ Department of Physics, University of Oslo, Oslo, Norway\\
$^{20}$ Department of Physics and Technology, University of Bergen, Bergen, Norway\\
$^{21}$ Dipartimento di Fisica, Universit\`{a} di Pavia, Pavia, Italy\\
$^{22}$ Dipartimento di Fisica dell'Universit\`{a} and Sezione INFN, Cagliari, Italy\\
$^{23}$ Dipartimento di Fisica dell'Universit\`{a} and Sezione INFN, Trieste, Italy\\
$^{24}$ Dipartimento di Fisica dell'Universit\`{a} and Sezione INFN, Turin, Italy\\
$^{25}$ Dipartimento di Fisica e Astronomia dell'Universit\`{a} and Sezione INFN, Bologna, Italy\\
$^{26}$ Dipartimento di Fisica e Astronomia dell'Universit\`{a} and Sezione INFN, Catania, Italy\\
$^{27}$ Dipartimento di Fisica e Astronomia dell'Universit\`{a} and Sezione INFN, Padova, Italy\\
$^{28}$ Dipartimento di Fisica `E.R.~Caianiello' dell'Universit\`{a} and Gruppo Collegato INFN, Salerno, Italy\\
$^{29}$ Dipartimento DISAT del Politecnico and Sezione INFN, Turin, Italy\\
$^{30}$ Dipartimento di Scienze MIFT, Universit\`{a} di Messina, Messina, Italy\\
$^{31}$ Dipartimento Interateneo di Fisica `M.~Merlin' and Sezione INFN, Bari, Italy\\
$^{32}$ European Organization for Nuclear Research (CERN), Geneva, Switzerland\\
$^{33}$ Faculty of Electrical Engineering, Mechanical Engineering and Naval Architecture, University of Split, Split, Croatia\\
$^{34}$ Faculty of Nuclear Sciences and Physical Engineering, Czech Technical University in Prague, Prague, Czech Republic\\
$^{35}$ Faculty of Physics, Sofia University, Sofia, Bulgaria\\
$^{36}$ Faculty of Science, P.J.~\v{S}af\'{a}rik University, Ko\v{s}ice, Slovak Republic\\
$^{37}$ Faculty of Technology, Environmental and Social Sciences, Bergen, Norway\\
$^{38}$ Frankfurt Institute for Advanced Studies, Johann Wolfgang Goethe-Universit\"{a}t Frankfurt, Frankfurt, Germany\\
$^{39}$ Fudan University, Shanghai, China\\
$^{40}$ Gauhati University, Department of Physics, Guwahati, India\\
$^{41}$ Helmholtz-Institut f\"{u}r Strahlen- und Kernphysik, Rheinische Friedrich-Wilhelms-Universit\"{a}t Bonn, Bonn, Germany\\
$^{42}$ Helsinki Institute of Physics (HIP), Helsinki, Finland\\
$^{43}$ High Energy Physics Group,  Universidad Aut\'{o}noma de Puebla, Puebla, Mexico\\
$^{44}$ Horia Hulubei National Institute of Physics and Nuclear Engineering, Bucharest, Romania\\
$^{45}$ HUN-REN Wigner Research Centre for Physics, Budapest, Hungary\\
$^{46}$ Indian Institute of Technology Bombay (IIT), Mumbai, India\\
$^{47}$ Indian Institute of Technology Indore, Indore, India\\
$^{48}$ INFN, Laboratori Nazionali di Frascati, Frascati, Italy\\
$^{49}$ INFN, Sezione di Bari, Bari, Italy\\
$^{50}$ INFN, Sezione di Bologna, Bologna, Italy\\
$^{51}$ INFN, Sezione di Cagliari, Cagliari, Italy\\
$^{52}$ INFN, Sezione di Catania, Catania, Italy\\
$^{53}$ INFN, Sezione di Padova, Padova, Italy\\
$^{54}$ INFN, Sezione di Pavia, Pavia, Italy\\
$^{55}$ INFN, Sezione di Torino, Turin, Italy\\
$^{56}$ INFN, Sezione di Trieste, Trieste, Italy\\
$^{57}$ Inha University, Incheon, Republic of Korea\\
$^{58}$ Institute for Gravitational and Subatomic Physics (GRASP), Utrecht University/Nikhef, Utrecht, Netherlands\\
$^{59}$ Institute of Experimental Physics, Slovak Academy of Sciences, Ko\v{s}ice, Slovak Republic\\
$^{60}$ Institute of Physics, Homi Bhabha National Institute, Bhubaneswar, India\\
$^{61}$ Institute of Physics of the Czech Academy of Sciences, Prague, Czech Republic\\
$^{62}$ Institute of Space Science (ISS), Bucharest, Romania\\
$^{63}$ Institut f\"{u}r Kernphysik, Johann Wolfgang Goethe-Universit\"{a}t Frankfurt, Frankfurt, Germany\\
$^{64}$ Instituto de Ciencias Nucleares, Universidad Nacional Aut\'{o}noma de M\'{e}xico, Mexico City, Mexico\\
$^{65}$ Instituto de F\'{i}sica, Universidade Federal do Rio Grande do Sul (UFRGS), Porto Alegre, Brazil\\
$^{66}$ Instituto de F\'{\i}sica, Universidad Nacional Aut\'{o}noma de M\'{e}xico, Mexico City, Mexico\\
$^{67}$ iThemba LABS, National Research Foundation, Somerset West, South Africa\\
$^{68}$ Jeonbuk National University, Jeonju, Republic of Korea\\
$^{69}$ Korea Institute of Science and Technology Information, Daejeon, Republic of Korea\\
$^{70}$ Laboratoire de Physique Subatomique et de Cosmologie, Universit\'{e} Grenoble-Alpes, CNRS-IN2P3, Grenoble, France\\
$^{71}$ Lawrence Berkeley National Laboratory, Berkeley, California, United States\\
$^{72}$ Lund University Department of Physics, Division of Particle Physics, Lund, Sweden\\
$^{73}$ Marietta Blau Institute, Vienna, Austria\\
$^{74}$ Nagasaki Institute of Applied Science, Nagasaki, Japan\\
$^{75}$ Nara Women{'}s University (NWU), Nara, Japan\\
$^{76}$ National Centre for Nuclear Research, Warsaw, Poland\\
$^{77}$ National Institute of Science Education and Research, Homi Bhabha National Institute, Jatni, India\\
$^{78}$ National Nuclear Research Center, Baku, Azerbaijan\\
$^{79}$ National Research and Innovation Agency - BRIN, Jakarta, Indonesia\\
$^{80}$ Niels Bohr Institute, University of Copenhagen, Copenhagen, Denmark\\
$^{81}$ Nikhef, National institute for subatomic physics, Amsterdam, Netherlands\\
$^{82}$ Nuclear Physics Group, STFC Daresbury Laboratory, Daresbury, United Kingdom\\
$^{83}$ Nuclear Physics Institute of the Czech Academy of Sciences, Husinec-\v{R}e\v{z}, Czech Republic\\
$^{84}$ Oak Ridge National Laboratory, Oak Ridge, Tennessee, United States\\
$^{85}$ Ohio State University, Columbus, Ohio, United States\\
$^{86}$ Physics department, Faculty of science, University of Zagreb, Zagreb, Croatia\\
$^{87}$ Physics Department, Panjab University, Chandigarh, India\\
$^{88}$ Physics Department, University of Jammu, Jammu, India\\
$^{89}$ Physics Program and International Institute for Sustainability with Knotted Chiral Meta Matter (WPI-SKCM$^{2}$), Hiroshima University, Hiroshima, Japan\\
$^{90}$ Physikalisches Institut, Eberhard-Karls-Universit\"{a}t T\"{u}bingen, T\"{u}bingen, Germany\\
$^{91}$ Physikalisches Institut, Ruprecht-Karls-Universit\"{a}t Heidelberg, Heidelberg, Germany\\
$^{92}$ Physik Department, Technische Universit\"{a}t M\"{u}nchen, Munich, Germany\\
$^{93}$ Politecnico di Bari and Sezione INFN, Bari, Italy\\
$^{94}$ Research Division and ExtreMe Matter Institute EMMI, GSI Helmholtzzentrum f\"ur Schwerionenforschung GmbH, Darmstadt, Germany\\
$^{95}$ Saga University, Saga, Japan\\
$^{96}$ Saha Institute of Nuclear Physics, Homi Bhabha National Institute, Kolkata, India\\
$^{97}$ School of Physics and Astronomy, University of Birmingham, Birmingham, United Kingdom\\
$^{98}$ Secci\'{o}n F\'{\i}sica, Departamento de Ciencias, Pontificia Universidad Cat\'{o}lica del Per\'{u}, Lima, Peru\\
$^{99}$ SUBATECH, IMT Atlantique, Nantes Universit\'{e}, CNRS-IN2P3, Nantes, France\\
$^{100}$ Sungkyunkwan University, Suwon City, Republic of Korea\\
$^{101}$ Suranaree University of Technology, Nakhon Ratchasima, Thailand\\
$^{102}$ Technical University of Ko\v{s}ice, Ko\v{s}ice, Slovak Republic\\
$^{103}$ The Henryk Niewodniczanski Institute of Nuclear Physics, Polish Academy of Sciences, Cracow, Poland\\
$^{104}$ The University of Texas at Austin, Austin, Texas, United States\\
$^{105}$ Universidad Aut\'{o}noma de Sinaloa, Culiac\'{a}n, Mexico\\
$^{106}$ Universidade de S\~{a}o Paulo (USP), S\~{a}o Paulo, Brazil\\
$^{107}$ Universidade Estadual de Campinas (UNICAMP), Campinas, Brazil\\
$^{108}$ Universidade Federal do ABC, Santo Andre, Brazil\\
$^{109}$ Universitatea Nationala de Stiinta si Tehnologie Politehnica Bucuresti, Bucharest, Romania\\
$^{110}$ University of Cape Town, Cape Town, South Africa\\
$^{111}$ University of Derby, Derby, United Kingdom\\
$^{112}$ University of Houston, Houston, Texas, United States\\
$^{113}$ University of Jyv\"{a}skyl\"{a}, Jyv\"{a}skyl\"{a}, Finland\\
$^{114}$ University of Kansas, Lawrence, Kansas, United States\\
$^{115}$ University of Liverpool, Liverpool, United Kingdom\\
$^{116}$ University of Science and Technology of China, Hefei, China\\
$^{117}$ University of Silesia in Katowice, Katowice, Poland\\
$^{118}$ University of South-Eastern Norway, Kongsberg, Norway\\
$^{119}$ University of Tennessee, Knoxville, Tennessee, United States\\
$^{120}$ University of the Witwatersrand, Johannesburg, South Africa\\
$^{121}$ University of Tokyo, Tokyo, Japan\\
$^{122}$ University of Tsukuba, Tsukuba, Japan\\
$^{123}$ Universit\"{a}t M\"{u}nster, Institut f\"{u}r Kernphysik, M\"{u}nster, Germany\\
$^{124}$ Universit\'{e} Clermont Auvergne, CNRS/IN2P3, LPC, Clermont-Ferrand, France\\
$^{125}$ Universit\'{e} de Lyon, CNRS/IN2P3, Institut de Physique des 2 Infinis de Lyon, Lyon, France\\
$^{126}$ Universit\'{e} de Strasbourg, CNRS, IPHC UMR 7178, F-67000 Strasbourg, France, Strasbourg, France\\
$^{127}$ Universit\'{e} Paris-Saclay, Centre d'Etudes de Saclay (CEA), IRFU, D\'{e}partment de Physique Nucl\'{e}aire (DPhN), Saclay, France\\
$^{128}$ Universit\'{e}  Paris-Saclay, CNRS/IN2P3, IJCLab, Orsay, France\\
$^{129}$ Universit\`{a} degli Studi di Foggia, Foggia, Italy\\
$^{130}$ Universit\`{a} del Piemonte Orientale, Vercelli, Italy\\
$^{131}$ Universit\`{a} di Brescia, Brescia, Italy\\
$^{132}$ Variable Energy Cyclotron Centre, Homi Bhabha National Institute, Kolkata, India\\
$^{133}$ Warsaw University of Technology, Warsaw, Poland\\
$^{134}$ Wayne State University, Detroit, Michigan, United States\\
$^{135}$ Yale University, New Haven, Connecticut, United States\\
$^{136}$ Yildiz Technical University, Istanbul, Turkey\\
$^{137}$ Yonsei University, Seoul, Republic of Korea\\
$^{138}$ Affiliated with an institute formerly covered by a cooperation agreement with CERN\\
$^{139}$ Affiliated with an international laboratory covered by a cooperation agreement with CERN.\\

\end{flushleft} 

\end{document}